\documentclass[9pt,twocolumn,twoside]{osajnl}

\usepackage{amsmath}
\usepackage{amsfonts}
\usepackage{amssymb}
\usepackage{amsthm}
\usepackage{bbm} 
\usepackage{bm}  

\journal{josaa} 

\setboolean{shortarticle}{false} 

\title{Ray and Wave Aberrations Revisited: A Huygens\reply{-Like} Construction \reply{Y}ields Exact Relations}

\author[1,2]{John Restrepo}
\author[3]{Pawel  J. Stoerck}
\author[1,2,*]{Ivo Ihrke}
\affil[1]{INRIA,Talence, France}
\affil[2]{LP2N, Talence, France}
\affil[3]{independent scholar, Isle St. Georges, France}
\affil[*]{Corresponding author: ivo.ihrke@inria.com}

\dates{Compiled \today}

\ociscodes{(220.1010) Aberrations, (080.0080) Geometric Optics}

\doi{\url{http://dx.doi.org/10.1364/ao.XX.XXXXXX}}

\newcommand{\comm}[1]{}

\newcommand{\reply}[1]{{#1}}

\newcommand{\wavefront}{\ensuremath{\boldsymbol{\varphi}}}
\newcommand{\wfa}{\ensuremath{\boldsymbol{\psi}}}

\newcommand{\trax}{{\epsilon_x}}
\newcommand{\tray}{{\epsilon_y}}

\newcommand{\fieldR}{\mathbb{R}}

\newcommand{\point}[1]{{\mathbf{#1}}}

\newcommand{\p}{\point{p}}
\newcommand{\s}{\point{s}}
\renewcommand{\i}{\point{i}}
\newcommand{\q}{\point{q}}

\newcommand{\x}{x}
\newcommand{\y}{y}
\newcommand{\z}{z}

\newcommand{\xp}{{\x_p}}
\newcommand{\yp}{{\y_p}}

\newcommand{\xs}{{\x_s}}
\newcommand{\ys}{{\y_s}}
\newcommand{\zs}{{\z_s}}

\newcommand{\R}{R} 
\newcommand{\W}{W} 
\newcommand{\n}{\point{n}} 

\newcommand{\ch}{\hat{\point{c}}}
\newcommand{\ph}{\hat{\p}}
\newcommand{\qh}{\hat{\q}}

\newcommand{\xh}{\hat{x}}

\newcommand{\Rh}{\hat{\R}}
\newcommand{\Wh}{\hat{\W}}
\newcommand{\nh}{\hat{\n}}
\newcommand{\phih}{\hat{\wavefront}}

\newcommand{\chxh}{\hat{\point{c}}_{\hat{X}}}
\newcommand{\chyh}{\hat{\point{c}}_{\hat{Y}}}

\newcommand{\nf}{n_f}

\newcommand{\Tpinv}{{\mathbf{T}_{\p(\xp)}^{-1}}}
\newcommand{\Tpinvel}[1]{{T_{#1}^{-1}}}

\newcommand{\dWdx}{\frac{\partial \W}{\partial \xp}}
\newcommand{\dWdy}{\frac{\partial \W}{\partial \yp}}

\newcommand{\dnfdx}{\frac{\partial \nf}{\partial \xp}}

\newcommand{\homzerovec}{{\left(\begin{array}{c}0\\0\\1\end{array}\right)}}

\newcommand{\yh}{\hat{\y}}
\renewcommand{\Wh}{{\hat{\W}}}

\renewcommand{\Tpinv}{{T_\p^{-1}}}

\newcommand{\grad}[1]{{\nabla{#1}}}
\newcommand{\gradW}{\grad{\W}}
\newcommand{\gradWhat}{\grad{\Wh}}

\newcommand{\Tbar}{\bar{\mathcal{T}}}

\newcommand{\Tbartransp}{\displaystyle{\Tbar^T}}
\newcommand{\dTbardx}{\frac{\partial \Tbar}{\partial \xp}}

\newcommand{\opdtoip}{t}

\newcommand{\mm}{{\textrm{mm}}}
\newcommand{\nm}{{\textrm{nm}}}

\theoremstyle{definition}
\newtheorem*{defn}{Definition}

\begin{abstract}

  The aberrations of a\reply{n optical} system can be described in terms of the
  wave aberrations, defined as the departure from the ideal spherical
  wavefront; or the ray aberrations, which are in turn the deviations
  from the paraxial ray intersection\reply{s} measured in the image plane. The
  classical connection between the two descriptions is an
  approximation, the error of which has, so far, not been quantified
  analytically.

  We derive exact analytical equations for computing the wavefront
  surface, the aberrated ray directions, and the transverse ray
  aberrations in terms of the wave aberrations (OPD) and the reference
  sphere.  We introduce precise conditions for a function to be an OPD
  function, show that every such function has an associated wavefront,
  and study the error arising from the classical approximation. We
  establish strict conditions for the error to be small. We illustrate
  our results with numerical simulations.  Our results show that large
  numerical apertures and OPD functions \reply{with strong gradients}
  yield larger approximation errors.
  
\end{abstract}

\setboolean{displaycopyright}{true}

\begin{document}

\maketitle
\thispagestyle{fancy}
\ifthenelse{\boolean{shortarticle}}{\abscontent}{}

\section{Introduction}\label{sec:intro}
\vspace{-0.1cm}
The geometrical theory of aberrations adjusts the predictions of
paraxial optics to a more realistic depiction of how a real lens
performs. Two of its descriptors are the ray aberrations and the wave
aberrations. Both concepts are directly related\reply{,} as has been shown in
the classic literature~\cite{ob:bornwolf, ob:welford, ob:mahajan,
  ob:malacara}. The ray aberration\reply{s describe} the deviation between the
aberrated rays and the paraxial/ideal rays as a \reply{transverse} distance
measured in the image plane.  The wave aberration describes the
deviation of the aberrated wavefront as compared to an ideal spherical
wavefront that produces a perfect image point. Equivalently, one can
consider the wave aberrations as the differences in time of flight of
the light along an aberrated ray with respect to the time it would
take to reach the image along a paraxial ray, hence the alternative
use of the name optical path differences (OPD).

Both concepts, the ray aberrations and wave aberrations are commonly
related by means of \cite{ob:bornwolf, ob:wyantcreath}

\begin{eqnarray}
\frac{\partial \W}{\partial \x} &\approx& - \frac{\trax}{r}, \,\, \textrm{and} \nonumber \\
\frac{\partial \W}{\partial \y} &\approx& - \frac{\tray}{r},
\label{eq:classic_eq}
\end{eqnarray}

where $W$ represents the \reply{wave} aberration (OPD), $\trax$ and
$\tray$ are the ray aberrations, and $r$ is the radius of the ideal
wavefront\reply{,} also known as the reference sphere.

The approximation in Eqs.~(\ref{eq:classic_eq}) is commonly held to be
valid for small numerical apertures and small aberrations.  In spite
of this, they are widely used for instance in the analysis of the
Hartmann-Shack sensor~\cite{oa:platt2001,Neal2002} or in optical
design software.

While exact relations between quantities indirectly related to the
aberrations~\cite{Mejia2012726}, or based on re-definitions of the
optical path difference~\cite{oa:rayces1964}, have appeared in the
literature, there is, so far, no such relation for the standard
definition of the optical path difference as a phase difference along
the aberrated ray.

In this paper, we derive exact analytical equations to compute
wavefront points, aberrated ray directions and the transverse ray
aberrations in terms of the OPD function along the aberrated
ray. Whereas the classical equations, Eqs.~(\ref{eq:classic_eq}), are
only an approximation, the new equations are applicable to large
numerical aperture settings and for arbitrary differentiable OPD
functions. We prove the exactness of the equations by validating the
defining properties of the wavefront, i.e. the distance to the
reference sphere and the orthogonality with the aberrated rays. We
show that the classical equations for the ray aberrations are a
special case of our equations and detail the conditions for a good
approximation. Finally, we evaluate the approximation error
quantitatively.

\section{Overview}

The main tool for our derivation is a Huygens-like interpretation of
the wavefront as an envelope of spheres with a varying radius that is
given by the OPD function\reply{, Fig.~\ref{fig:opd_norm_smaller_one}a.}

This conception enables us to perform limit considerations that are
most suitably studied in the tangent space of the reference sphere.
The derivations are initially performed in this local space rather
than in exit pupil coordinates. In the new coordinate system, we
arrive at exact analytic equations for the wavefront and the aberrated
ray directions, which, by means of a suitable transformation can be
related to the original exit pupil coordinates. In this scheme, the
aberrated ray directions can be computed without differentiating the
wavefront.

As a result, we obtain a set of rays with origins at the wavefront
that can be propagated to the image plane to compute the exact
analytic expressions for the transverse ray aberrations.

We perform the derivation in several steps. First,
Sect.~\ref{sec:1D_derivation}, we perform the basic geometric
construction of a wavefront tangent in one dimension in the canonical
setting afforded by the tangent space construction outlined above.

We leave the detailed definition of the required coordinate
transformation for the discussion of the two-dimensional case,
Sect.~\ref{sec:coordinates}. We introduce the transformation between
global and local coordinate frames, putting special emphasis on the
transformation of functions defined in global exit pupil coordinates
to the local tangent frame systems.

We then generalize the one-dimensional geometric argument to two
dimensions, Sect.~\ref{sec:2D_derivation}, and derive local expressions
for the wavefront and the aberrated ray directions. We continue by
linking these expressions back to exit pupil coordinates, both in
their arguments and in their values. We arrive at the key results,
Eqs.~(\ref{eq:wfglob2D}) and ~(\ref{eq:normalglob2D}).

Finally, we make a connection to the transverse ray aberrations and
the classical approximation, Eqs.~(\ref{eq:classic_eq}), elucidating
the conditions for a valid approximation. Sect.~\ref{sec:examples}
demonstrates exemplary applications and quantitative properties of the
equations.

\reply{ The Appendix contains a proof of the wavefront properties of
  the derived quantities and establishes that every OPD function
  corresponds to a wavefront.}

\section{The Wave Aberration Function (OPD)}
\label{sec:alternative_definitions}

\subsection{OPD Definitions}

In the classic literature, there are two different recurrent
definitions for the \reply{wave aberration function (OPD)}.

The standard definition~\cite{ob:bornwolf,ob:wyantcreath,ob:mahajan}
considers the path length between points on the aberrated wavefront
and on the reference sphere, connected along the direction of the 
propagation of the aberrated ray. The OPD value is reported for the
coordinate \reply{of} the reference sphere point in the exit pupil plane. Its
relation to the ray aberration is stated in
Eqs.~(\ref{eq:classic_eq}). 

The alternative definition~\cite{oa:rayces1964,ob:mahajan} considers the wave
aberration to be measured along a radius of the reference sphere,
again between two points on the wavefront and on the reference sphere,
but with the OPD value assigned to the exit pupil location of the
wavefront point.

\begin{figure}[tbp]
\centering
\fbox{\includegraphics[width=0.85\linewidth]{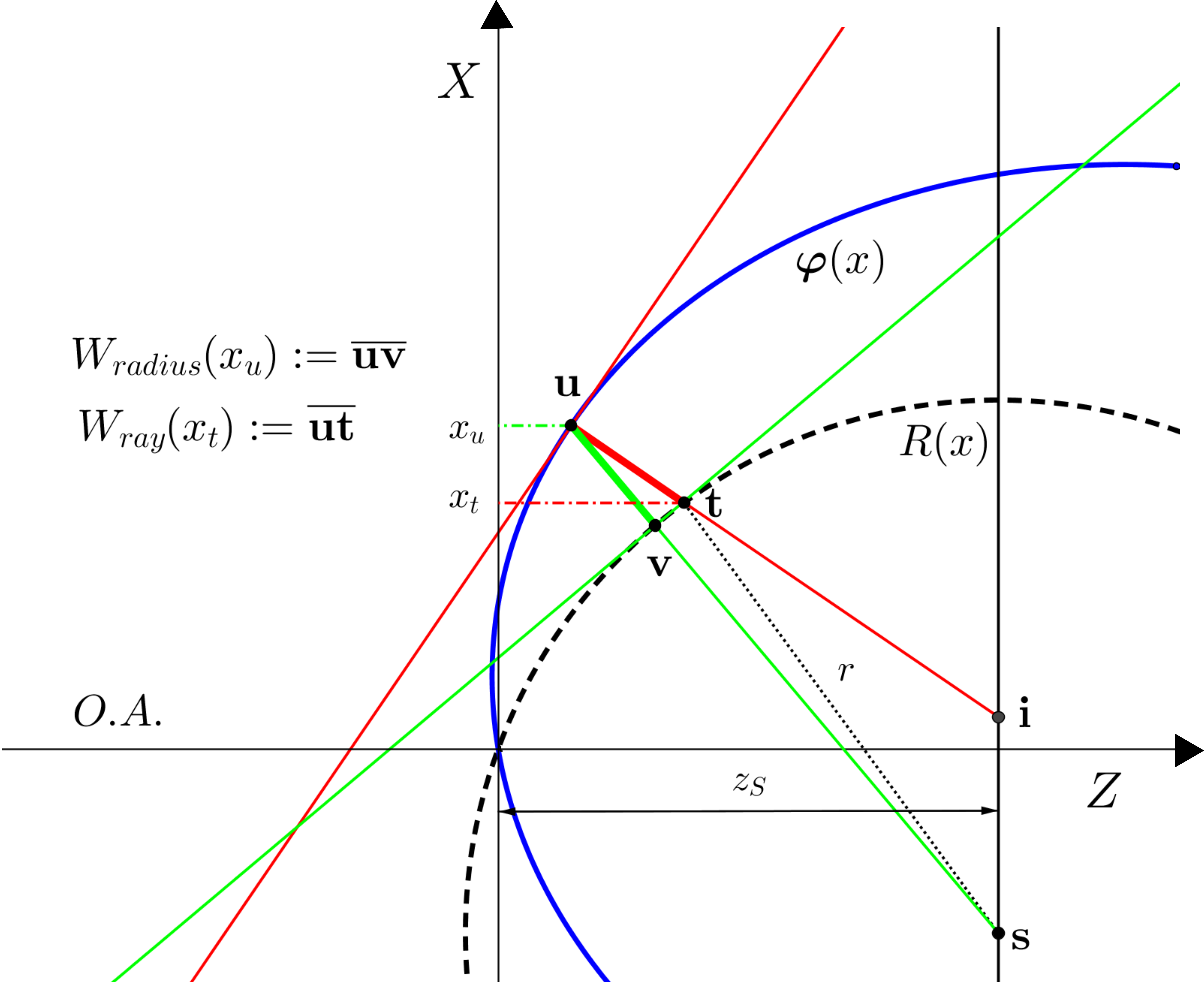}}
\caption{(Color online) Alternative definitions of the wave
  aberrations. The exit pupil is located along the $X$ axis, while the
  image plane is located at a distance $z_S$ along the $Z$ axis. The
  wavefront is the curve $\wavefront(x)$. The reference sphere $\R(x)$
  is centered at the paraxial image point $\s$ and has a radius of
  $r$. The aberrated ray passing through the point $\mathbf{u}$ intersects the
  image plane at the aberrated image point $\i$. The transverse ray
  aberration is the difference $\i-\s$.}
\label{fig:OPD_definitions}
\end{figure}

Both definitions are illustrated in Fig. \ref{fig:OPD_definitions} for
a point $\mathbf{u}$ on the wavefront. For simplicity, we refer to the first
definition as $\W_{ray}$ since the path lengths are measured along the
ray, while the second one is denoted as $\W_{radius}$. The pupil
coordinate $\x_{t}$ represents the pupil coordinate position for
$\W_{ray}$, while $\x_{u}$ is the same for the alternative wave
aberration definition. The wave aberration value is given a sign that
depends on the delay relation between the wavefront and the reference
sphere. For the particular case of point $\mathbf{u}$ in
Fig.~\ref{fig:OPD_definitions}, the wavefront is delayed which implies
a negative sign of the OPD.

For $\W_{radius}$ the resulting relation between ray and wave
aberration is different from Eqs.~(\ref{eq:classic_eq}) and is
reported~\cite{oa:rayces1964} as:

\begin{eqnarray}
  \frac{\partial \W_{radius}}{\partial \x} &=& -\frac{\trax}{r - \W_{radius}}, \,\, \textrm{and} \nonumber \\
  \frac{\partial \W_{radius}}{\partial \y} &=& -\frac{\tray}{r - \W_{radius}}. 
  \label{eq:classic_rayces_eq}
\end{eqnarray}

The advantage of the $\W_{radius}$ definition is a simple computation of the wavefront surface via triangle
relationships. However, Eqs.~(\ref{eq:classic_rayces_eq}) are
differential equations as compared to the classical
Eqs.~(\ref{eq:classic_eq}) and therefore, even though they are
exact~\cite{oa:rayces1964}, difficult to solve.

On the other hand, for the standard definition, $\W_{ray}$, there is no
obvious construction of the wavefront surface from the OPD values,
since the aberrated ray directions are unknown.

Most authors prefer the $\W_{ray}$ definition since it provides a
direct connection with the pupil function and the calculation of the
point spread function (PSF)~\cite{ob:welford}, but often the
distinction is not clearly made.

In this article, we derive exact equations for the wavefront and the
aberrated ray directions for the standard OPD definition $\W_{ray}$ and
use them to make an exact link with the ray aberrations.

\subsection{Properties of the OPD Function}

Since the derivations critically depend on the exact properties of the
OPD definition, we discuss the interpretation underlying our
derivations in detail.

We use the standard definition of the OPD function $W_{ray}$ as given
above, simply denoting it as $W$ in the following. In contrast to most
of the literature, we interpret the domain of the OPD function to be
the reference sphere. The usual parameterization in terms of exit
pupil coordinates is, in this sense, one parameterization of the
function's domain. Other parameterizations are possible and we will
use this insight to define the OPD function in local coordinate
systems, Sect.~\ref{sec:coordinates}. These differently parameterized
OPD functions all describe the same quantity, i.e. the phase delay
along an aberrated ray intersecting the reference sphere at the
position of intersection, just with a different frame of reference.

\begin{figure}[t]
  \includegraphics[width=\columnwidth]{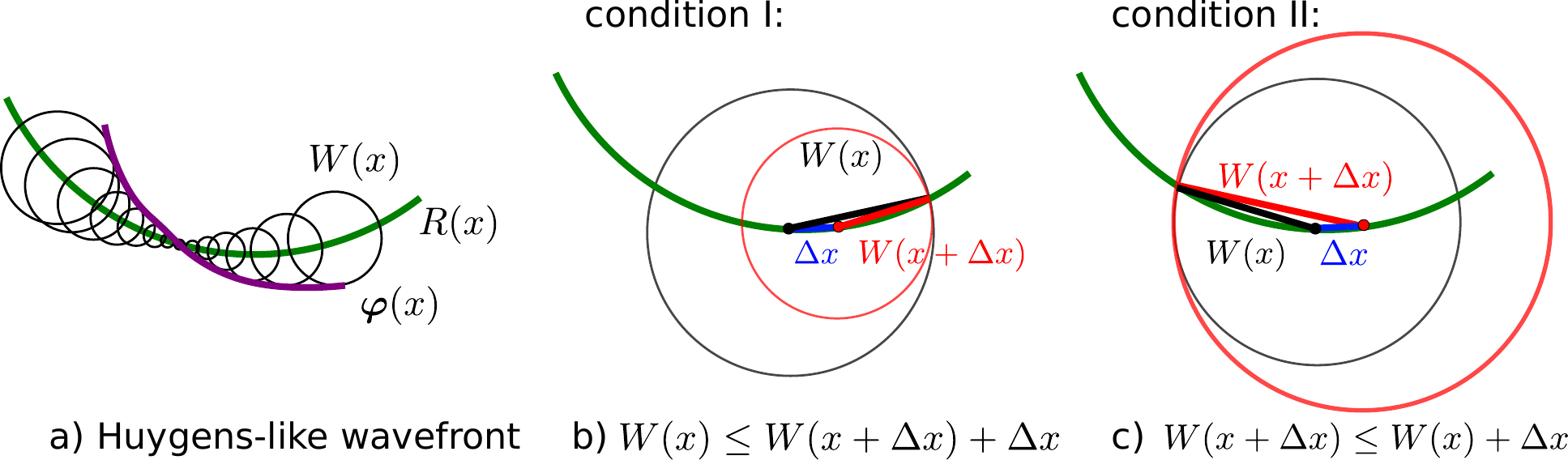}
  \caption{(Color online) \reply{ a) Huygens-like wavefront
      construction illustrated - the wavefront $\wavefront(x)$ is the
      envelope of spheres with radii defined by the OPD function
      $\W(x)$.} b) and c) Illustrating the condition that the gradient
    norm of the OPD function is smaller or equal to one. Applying a
    Taylor expansion to the term $\W(\x+\Delta \x)$ and simplifying
    the triangle inequality yields the condition. \reply{The colors of
      the mathematical terms correspond to the colored segments
      indicating the distances that constitute the triangle.} }
  \label{fig:opd_norm_smaller_one}
\end{figure}

A direct consequence of the above considerations and the fact that the
OPD is single-valued is that no wavefront point can be at a closer
distance to the reference sphere than the OPD value. This implies that
there is an open ball around any point on the reference sphere that
does not contain wavefront points. The radius of this open ball is
equal to the OPD value. A sphere with this radius, i.e. the closure of
the open ball, is tangent to the wavefront. The wavefront can
therefore be considered as the envelope of a set of spheres with a
varying radius that is described by the OPD function, a statement of
Huygens principle. \reply{The concept is illustrated in
  Fig.~\ref{fig:opd_norm_smaller_one} a), where $\R(x)$ is the
  reference sphere, $\W(x)$ the OPD function, and $\wavefront(x)$ the
  resulting wavefront.}

A \reply{misconception} that is often found in the literature is an
ambiguity between the slope of the wavefront and the slope of the OPD
function. We emphasize that these two concepts must be distinguished.

A key property of an OPD function is that its gradient magnitude
cannot exceed one. To appreciate this point, consider a 1D setting,
Fig.~\ref{fig:opd_norm_smaller_one} b) and c).  As mentioned above,
the OPD value $W(x)$ implies an open ball without wavefront points
surrounding a particular point $x$ on the reference sphere. If the
norm of the OPD derivative is greater than one $|d\W/d\x| > 1$, the
radius of this ball changes more quickly than the evaluation position \reply{, i.e.}
$|d\W|>|d\x|$. It follows that one of the balls completely contains the
other -- which is a contradiction since both balls, by definition, do
not contain wavefront points, but are simultaneously tangent to it. It
follows that the norm of the OPD derivative cannot exceed one.  An
intuitive interpretation of this property is that the OPD would be
required to be multi-valued in this situation. An alternative
interpretation is that the wavefront can only be represented as an
envelope of balls if the condition on the OPD norm is satisfied.

\reply{The illustration in Fig.~\ref{fig:opd_norm_smaller_one} b) and
  c) shows the two cases for a finite displacement $\Delta x$. The
  constraint that one ball does not contain the other yields the
  triangle inequalities illustrated in the Figure. Passing to the
  limit as $\Delta x \to 0$ yields the condition $|d\W/d\x|\leq1$.}

\reply{With these prerequisites, we introduce the following }

\begin{defn}{{\bf OPD function:}}
  An OPD function $\W:S^2 \to \fieldR$ is a twice-differentiable
  function with a gradient norm smaller or equal to one. Its domain is
  the reference sphere.
\end{defn}

The existence of the second derivatives is a technical requirement for
subsequent developments. As implied by the definition, the gradient is
to be taken on the reference sphere.

In contrast to the above discussion, the \reply{\emph{wavefront}} derivative can have
arbitrary values. It follows that there are wavefronts that cannot be
represented by an OPD function. However, we show in the
\reply{Appendix} that all OPD functions\reply{, satisfying the above
  conditions,} describe wavefronts by deriving explicit construction
rules. The specification of an OPD function is therefore a sufficient
condition for a wavefront to exist.

\begin{figure}[tbp]
\centering
\fbox{\includegraphics[width=0.95\linewidth]{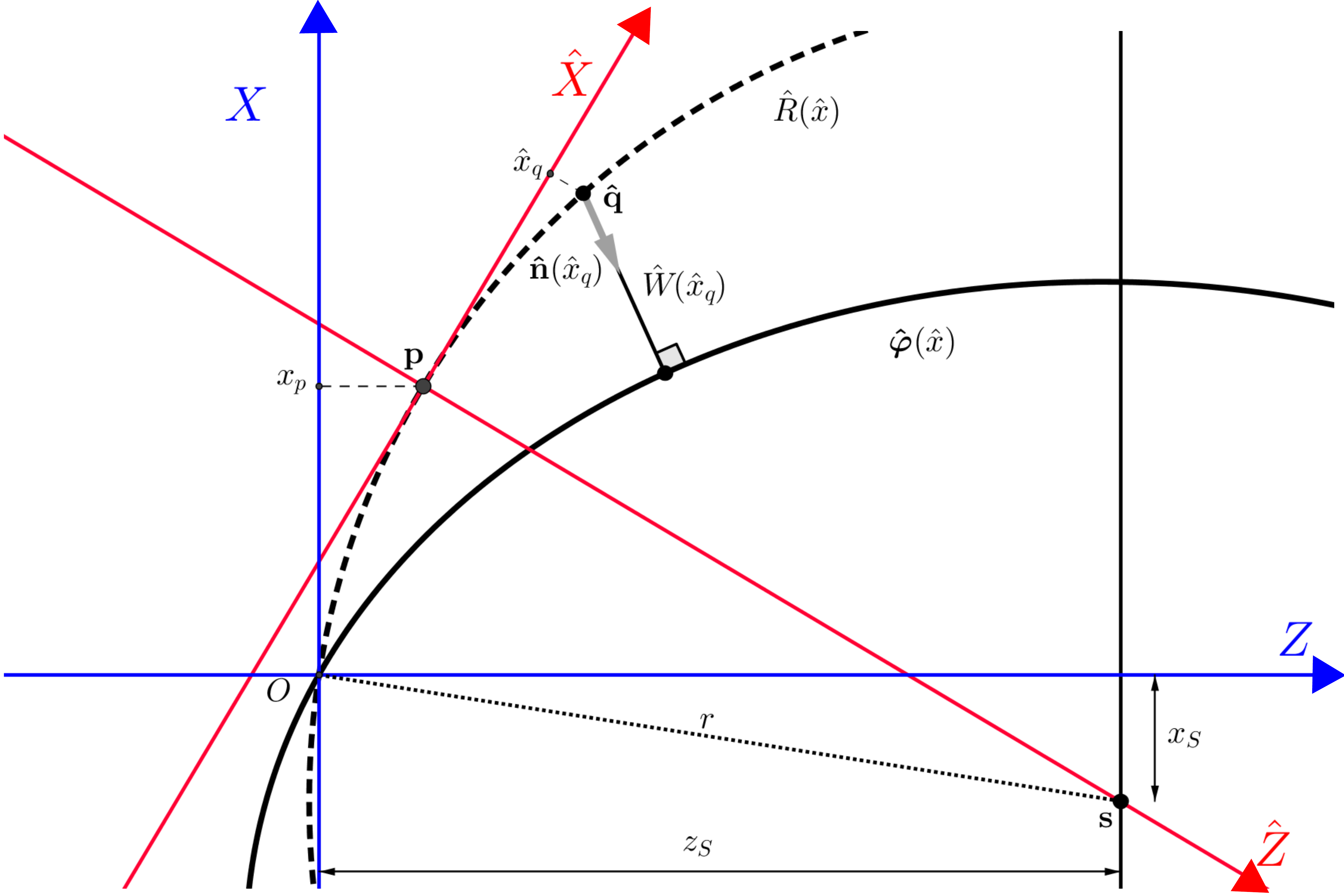}}
\caption{(Color online) Geometry considered. The axes $XZ$ represent
  the original coordinate system or \textit{global coordinates} and
  the axes $\hat{X}\hat{Z}$ represent the \textit{local coordinates}
  for the point $\p(\xp)$. We utilize circumflex symbols to distinguish
  quantities of the local system from those of the global one.}
\label{fig:axis_change}
\end{figure}

\section{Wavefront Points and Aberrated Ray Directions in a 1D Canonical Setting}
\label{sec:1D_derivation}

We develop the major geometric reasoning of our derivation in a
canonical one-dimensional setting. \reply{ Without loss of generality, 
we consider an orthonormal local coordinate system with
its origin on the reference sphere and spanning its tangent space.} To
complete the basis, we use the tangent plane normal, oriented towards
the paraxial image point $\s$. Such a local coordinate system is
depicted in Fig.~\ref{fig:axis_change} for the current
one-dimensional setting, for the point $\p$. In local coordinates
$\ph=(0,0)^T$, i.e. it coincides with the local coordinate origin.

As detailed in Sect.~\ref{sec:coordinates} for the full 2D case, these
local coordinate systems can be obtained by a rigid body
transformation of the exit pupil coordinate system.

All quantities relating to the local system are denoted with
circumflex symbols. In particular, $\ph$ is the local coordinate
origin, $\hat{X}$ and $\hat{Z}$ are the local coordinate axes and
$\hat{W}(\xh)$ is the OPD function parameterized in local coordinates
$\xh$.  The specifics of this parameterization are also covered in
Sect.~\ref{sec:coordinates} for the 2D-setting. In addition, we denote
the local representation of the reference sphere as $\Rh(\xh) := r -
(r^2-\xh^2)^{1/2}$ and the local aberrated ray directions as
$\nh(\xh)$. Given these quantities, we may express a wavefront point
as

\reply{
\begin{equation}
  \phih(\xh_{q}) = \qh(\xh_{q}) + \Wh(\xh_{q}) \cdot \nh(\xh_{q}),
  \label{eq:wavefrontpointlocal}
\end{equation}
}
where $\qh(\xh_q) = (\xh_q,\Rh(\xh_q))^T$ is a point on the reference
sphere corresponding to the local tangent space coordinate $\xh_q$,
see Fig.~\ref{fig:axis_change}. The local wavefront function $\phih:
\fieldR \to \fieldR^2$ yields the local 2D coordinates of the
wavefront point. Its horizontal coordinate may be different from the
evaluation position \reply{$\xh_{q}$}.

\subsection{Constructing a Tangent to the Wavefront}
\label{subsec:tan_circles}

Considering the fundamental definition of the wavefront as the surface
which is normal to the aberrated rays, we can consider a circle with
radius $r_1 = \Wh(0)$ that is centered in the point $\ph=(0,0)^T$,
i.e. in the origin of the local coordinate system. This circle must be
tangent to the wavefront since the wavefront is, by definition,
located at a distance $\Wh(0)$ from point $\ph$. We wish to determine
the point of intersection between the circle and the wavefront.

\begin{figure}
\centering
\fbox{\includegraphics[width=0.85\linewidth]{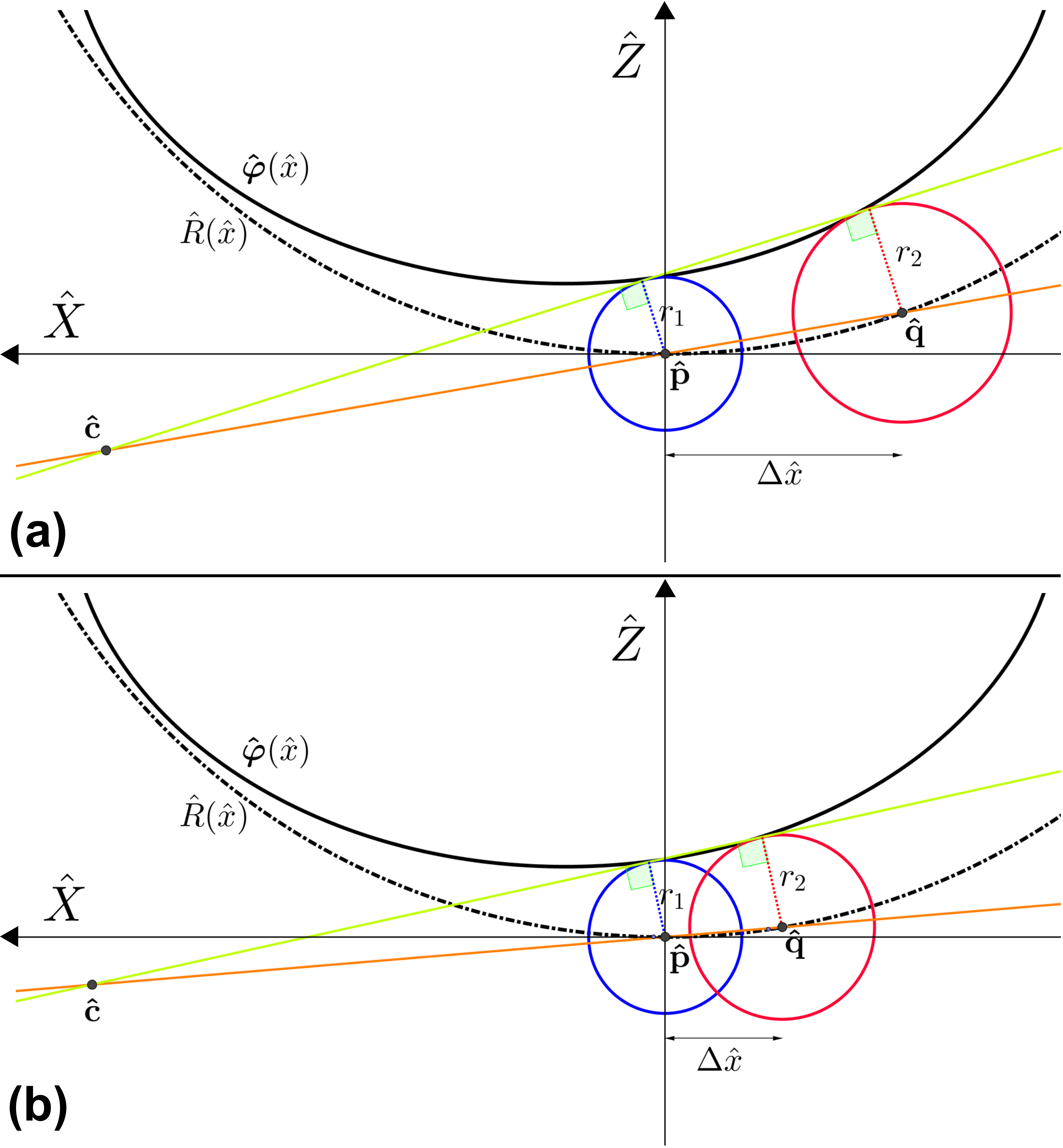}}
\caption{(Color online) Approaching circles. As point $\qh$ moves
  towards $\ph$ along the reference sphere, the circles get closer. The
  sub-images (a) and (b) represent a decrease in $\Delta\xh$.}
\label{fig:circles_and_tangent}
\end{figure}

For this, we introduce a neighboring second circle with its center at
point $\qh = (\xh_q,\Rh(\xh_q))^T$, also centered on the reference
sphere and also tangent to the wavefront with radius $r_2 =
\Wh(\xh_q)$. Both circles are shown in
Fig.~\ref{fig:circles_and_tangent}.

We now consider the circle at $\qh$ to be approaching the circle at
the center of the local coordinates $\ph$. In the limit, as the
horizontal distance $\Delta\xh = \xh_q$ between the centers tends
to zero, the two circles coincide and the tangent to both becomes the
tangent of the wavefront. Expressing this intuition mathematically
leads to equations for the ray direction and the wavefront itself.

Considering in more detail the geometry of
Fig.~\ref{fig:circles_and_tangent}, the tangent to both circles
intersects the line connecting their centers at the point $\ch$, where
distances between the points are related by the simple triangle
relation

\begin{equation}
\frac{r_{1}}{|\ph\ch|} = \frac{r_{2}}{|\qh\ch|}.
\end{equation}

Using \reply{$|\qh\ch|  = |\ph\ch| + |\ph\qh|$} we arrive at the following
equation

\reply{
\begin{equation}
|\ph\ch| = \frac{r_{1}|\ph\qh|}{r_{2} - r_{1}}. 
\end{equation}
}

The point $\ch$ is then given by:

\begin{equation}
  \ch = \ph - \frac{(\qh - \ph)r_{1}}{r_{2} - r_{1}}.
  \label{vectC}
\end{equation}

We now introduce the explicit expressions for the vector quantities

\begin{eqnarray}
\ph &=& (0,0)^T, \nonumber \\
\qh &=& (\Delta\xh, \Rh(\Delta\xh))^T = \left(\Delta\xh, \reply{r - (r^{2} - \Delta\xh^{2})^{1/2}}\right)^T, \nonumber \\
r_{1} &=& \Wh(0), \nonumber \\
r_{2} &=& \Wh(\Delta\xh) = \Wh(0) + \frac{d\Wh}{d\xh}\bigg|_{0}\Delta\xh + O(\Delta\xh^{2}),  
\end{eqnarray}

and insert the definitions into Eq.~(\ref{vectC}). We use a Taylor
expansion for $\Wh$ at the point $\qh$ for studying the limit as point
$\qh$ approaches point $\ph$. Since the OPD function $\Wh$ is
differentiable, the relation is exact in this limit. We thus obtain an
explicit equation for the vector

\begin{equation}
  \ch = -\frac{\Wh(0)
    \left(
    \begin{array}{c}
       \Delta\xh \\
        r - (r^{2} - \Delta\xh^{2})^{\reply{1/2}}
    \end{array}
    \right) }
         {\frac{d\Wh}{d\xh}|_{0}\Delta\xh + O(\Delta\xh^{2})},
   \label{vectC_definitions_inserted}
\end{equation}

which, as $\Delta \xh$ tends to zero becomes

\begin{equation}
  \lim_{\Delta\xh\to\\0} \ch =
  \left(
  \begin{array}{c}
    - \Wh(0) \bigg/ \frac{d\Wh}{d\xh}|_{0} \\ 0 \end{array} \right),
  \label{limitC}
\end{equation} 

\begin{figure}
\centering
\fbox{\includegraphics[width=0.85\linewidth]{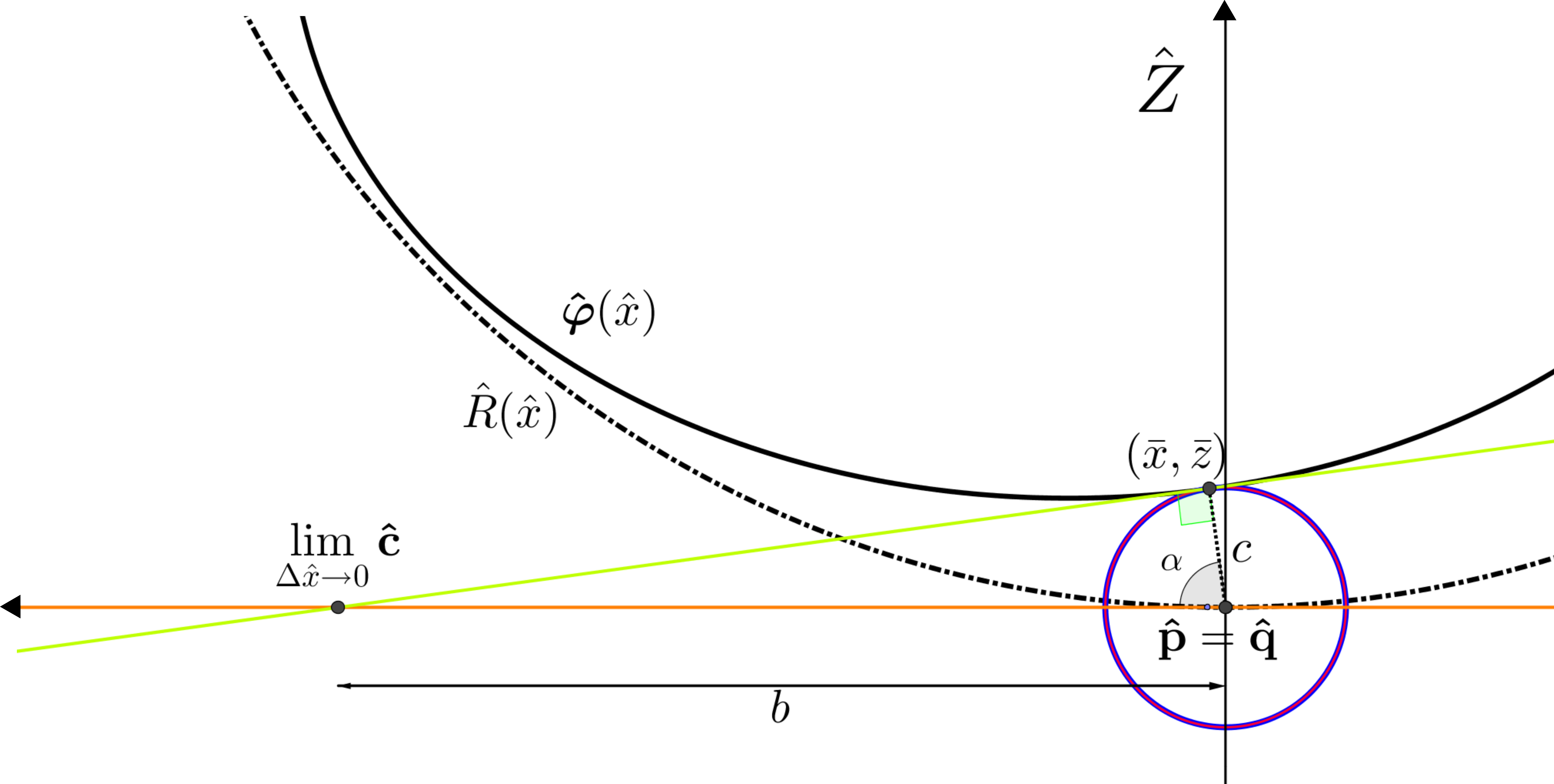}}
\caption{(Color online) Limit when circles coincide. The point $\ch$
  is now located on the axis.}
\label{fig:Tangents_coincide}
\end{figure}

where the limit has been determined using the rule of
$l'H\hat{o}pital$. The previous equation is illustrated in
Fig.~\ref{fig:Tangents_coincide}. The limit behavior of the point
$\ch$ is finite if $\frac{d\Wh}{d\xh}|_{0} \not= 0$ and it is located
on the $\hat{X}$ axis. For the case that $\frac{d\Wh}{d\xh}|_{0} = 0$,
the point $\ch$ is located at infinity. However the calculation of the
wavefront point is now trivial since it is located on the axis
$\hat{Z}$ at a distance $\Wh(0)$ from the origin of the local
coordinates.

For the limiting case where $\ch$ is finite, we determine the tangent
point with the wavefront utilizing the triangle between the tangent
point, the origin of the local coordinate system, and the limiting
point $\ch$. This triangle is illustrated in
Fig.~\ref{fig:Tangents_coincide}. Using the geometry, the tangential
point, with coordinates $(\bar{x},\bar{z} )^T$ is calculated as

\begin{eqnarray}
\bar{x} &=& c \times cos(\alpha), \nonumber \\
\bar{z} &=& (c^{2} - \bar{x}^{2})^{1/2}. 
\label{eq:xz_bar}
\end{eqnarray}

Trigonometry yields

\begin{equation}
\bar{x} = \frac{c^{2}}{b}.
\end{equation}

Note that there may be two solutions since two tangent lines to the
circles exist. However, since the sign of the OPD function indicates
the direction of the wavefront with respect to the reference sphere,
we can unambiguously select the correct solution.

\subsection{Wavefront Point and Aberrated Ray}
\label{subsec:wavefront_calcu}

Replacing the auxiliary variables with their definitions in terms of
the OPD function, 

\begin{eqnarray}
  c       &=& \Wh(0), \nonumber \\
  b       &=& -\frac{\Wh(0)}{\frac{d\Wh}{d\xh}|_{0}}, \nonumber \\
\end{eqnarray}

we arrive at explicit expressions for the local coordinates of the
wavefront point:

\begin{eqnarray}
  \bar{x} &=& -\Wh(0)\frac{d\Wh}{d\xh}\bigg|_{0}, \nonumber \\
  \bar{z} &=& \Wh(0) \left[ 1 - \left( \frac{d\Wh}{d\xh}\bigg|_{0}\right)^{2} \right] ^{1/2}.
\end{eqnarray}

We denote it in vector notation as $\phih(0)=(\bar{x},\bar{z})^T$:

\begin{equation}
  \phih(0) = \Wh(0)
  \left(
  \begin{array}{c}
    -\frac{d\Wh}{d\xh}\big|_{0} \\ 
    \left[ 1 - \left( \frac{d\Wh}{d\xh}\big|_{0}\right)^{2} \right] ^{1/2}
  \end{array}
  \right).
  \label{wavefront_at_local}
\end{equation}

Here we notice the significance of the condition that the norm of the
OPD derivative be smaller or equal to one,
Sect~\ref{sec:alternative_definitions}.

Eq.~(\ref{wavefront_at_local}) allows to simultaneously obtain the
normalized ray direction. As seen from the equation, the vectorial
part is a) a unit vector, and b) multiplied by the OPD value
$\Wh(0)$. Therefore, this vector component represents exactly the
normalized ray direction:

\begin{equation}
  \nh(0) =
  \left(
  \begin{array}{c}
    -\frac{d\Wh}{d\xh}\big|_{0} \\ 
    \left[ 1 - \left( \frac{d\Wh}{d\xh}\big|_{0}\right)^{2} \right] ^{1/2}
  \end{array}
  \right).
  \label{ray_direction_at_local}
\end{equation}

We interpret the result as showing that, in the canonical situation,
the OPD derivative equals the (negative) direction cosine of the ray.

Eqs.~(\ref{wavefront_at_local}) and~(\ref{ray_direction_at_local}) allow
us to obtain the exact wavefront point and aberrated ray direction
knowing only the local function for the reference sphere $\Rh(\xh)$ and
the local OPD function $\Wh(\xh)$.

\section{Local Coordinate Systems and Functions Therein}
\label{sec:coordinates}
\label{subsec:coord_transform}

So far, we have been describing the canonical situation in one
dimension. For a generalization of the result, we need to 1) detail
the construction of the local coordinate system as well as the
transfer of the relevant functions, and 2) expand the results of the
previous section towards two dimensions.

As mentioned in Sect.~\ref{sec:1D_derivation}, we require an
orthonormal tangent frame to the reference sphere and a transformation
$\Tpinv$ from this local system to the global exit pupil coordinate
system. The underlying reason for requiring orthonormality is a
preservation of distance measures in relation to the global coordinate
system.

The tangent frame of the reference sphere and its associated
transformation are parameterized by an evaluation point
$\p=(\xp,\yp,R(\xp,\yp))^T$ on it. The reference sphere in global exit
pupil coordinates is given by

\begin{equation}
  R(\xp,\yp) = \zs - (r^2 - (\xp - \xs)^2 -(\yp - \ys)^2)^{1/2}.
  \label{eq:refsphereglobal}
\end{equation}

The point $\s=(\xs,\ys,\zs)$ is the
paraxial image point, also in global coordinates, and $r=(\xs^2 +
  \ys^2 + \zs^2)^{1/2}$ is the radius of the reference sphere. The exit
pupil is centered in the global origin, and the positive $Z$-axis is
pointing towards the image plane, which is located at a distance
$\zs$.

\subsection{Local Orthonormal Tangent-Frame to the Reference Sphere}

We construct an orthonormal local tangent frame coordinate system at
point $\p$ by resorting to spherical coordinates around the paraxial
image point $\s$. Suitably normalized derivatives with respect to the
spherical coordinates then provide us with the local linear
approximation of the reference sphere, i.e. with its tangent plane. We
choose to position the poles of the spherical coordinate system in the
image plane, in particular along the $Y$-axis, in order to avoid
singularities of the reference sphere parameterization in the space
between the image plane and the exit pupil.

In particular, we use \reply{the following assignment of Euler angles}

\begin{eqnarray}
  \theta(\xp,\yp) &=& \tan^{-1} \left( \frac{R(\xp,\yp)-\zs}{\xp-\xs} \right),  \nonumber \\
  \phi(\xp,\yp) &=& \cos^{-1} \left( \frac{\yp-\ys}{r} \right),  \,\, \textrm{and}  \nonumber \\
  \rho(\xp,\yp) &=& \textrm{const.} = r,
  \label{eq:sphericalcoords}
\end{eqnarray}

the inverse equations of which are given by 

\begin{eqnarray}
  x &=& \rho \cos({\theta}) \sin({\phi}) + \xs, \nonumber \\
  y &=& \rho \cos({\phi}) + \ys, \,\, \textrm{and} \nonumber \\
  z &=& \rho \sin({\theta}) \sin({\phi}) + \zs.
\end{eqnarray}

The variables $\rho, \theta$ and $\phi$ all depend on the evaluation
position $(\xp,\yp)$ in global coordinates,
Eq.~(\ref{eq:sphericalcoords}).  The unit vectors of the tangent frame
are then obtained via

\begin{eqnarray}
  \vec{u}_x(\xp,\yp) &=& \frac{1}{\rho\sin{\phi}}\cdot
  \left(
  \begin{array}{c}
    \frac{\partial x}{\partial \theta}|_{(\xp,\yp)} \\
    \frac{\partial y}{\partial \theta}|_{(\xp,\yp)} \\
    \frac{\partial z}{\partial \theta}|_{(\xp,\yp)}
  \end{array}\right) =
  \left(\begin{array}{c}
    -\sin(\theta)\\
    0 \\
    \cos(\theta)
  \end{array}\right),   \nonumber \\
  \vec{u}_y(\xp,\yp) &=& \frac{1}{\rho}\cdot
  \left(
  \begin{array}{c}
    \frac{\partial x}{\partial \phi}|_{(\xp,\yp)} \\
    \frac{\partial y}{\partial \phi}|_{(\xp,\yp)} \\
    \frac{\partial z}{\partial \phi}|_{(\xp,\yp)}
  \end{array}\right) =
  \left(\begin{array}{c}
    \cos(\theta)\cos(\phi)\\
   -\sin(\phi) \\
    \sin(\theta)\cos(\phi)
  \end{array}\right),   \nonumber \\
  \vec{u}_z(\xp,\yp) &=&
  - \left(\begin{array}{c}
    \frac{\partial x}{\partial \rho}|_{(\xp,\yp)} \\
    \frac{\partial y}{\partial \rho}|_{(\xp,\yp)} \\
    \frac{\partial z}{\partial \rho}|_{(\xp,\yp)}
  \end{array}\right) =
  \left(\begin{array}{c}
   -\cos(\theta)\sin(\phi)\\
   -\cos(\phi) \\
   -\sin(\theta)\sin(\phi)
  \end{array}\right). \nonumber \\
\end{eqnarray}

Here, the vectors have been normalized and $\vec{u}_z$ has been
inverted to point towards the paraxial image point $\s$. Since the
unit vectors depend on $(\xp,\yp)$, the transformation from the local
coordinate systems to the global one is parameterized by the point of
evaluation $\p$. The associated matrix is given by

\reply{
\begin{eqnarray}
  \Tpinv &=&
   \left( \begin{array}{cccc}
    \vec{u}_x & \vec{u}_y & \vec{u}_z & \p \\
    0 & 0 & 0 & 1
   \end{array} \right), 
   \label{eq:Tpinv2D}
\end{eqnarray}
}
and its spatial derivatives are denoted as

\reply{
\begin{eqnarray}
  \frac{\partial \Tpinv}{\partial \xp} &=&
   \frac{\partial}{\partial \xp}\left( \begin{array}{cccc}
    \vec{u}_x & \vec{u}_y & \vec{u}_z & \p \\
    0 & 0 & 0 & 1
   \end{array} \right), 
   \label{eq:dTpinv2Ddx}
\end{eqnarray}
}


\reply{
and $  \frac{\partial \Tpinv}{\partial \yp}$ respectively.} We use homogeneous coordinates in order to describe
rigid body transformations, including their translational part, as
matrix-vector products.

\subsection{Transforming Exit Pupil Functions into Local Coordinates}
\label{subsec:coord_transform}

Key to the transfer of functions into the local coordinate systems is
the realization that they are defined with respect to a common surface
that is known in both systems. In particular, the reference sphere is
given in global coordinates by Eq.~\ref{eq:refsphereglobal}, whereas
in any local system it is

\begin{equation}
  \Rh(\xh,\yh) = r - ( r^2 - \xh^2 -\yh^2 )^{1/2},
\end{equation}

due to the symmetry of the sphere. Further, as mentioned in
Sect.~\ref{sec:alternative_definitions}, we define the domain of the
OPD function $\W$ as the surface of the reference sphere. The exit pupil
coordinate representation that is commonly used is then a
parameterization of this function on the sphere. Let us denote this
parameterization as $\W(\x,\y)$.

The OPD function value for a point $\q=(\x_q,\y_q,\R(\x_q,\y_q))^T$ on
the reference sphere is therefore obtained by evaluating
$\W(\x_q,\y_q)$.

\begin{figure}[tbp]
\centering
\fbox{\includegraphics[width=0.95\linewidth]{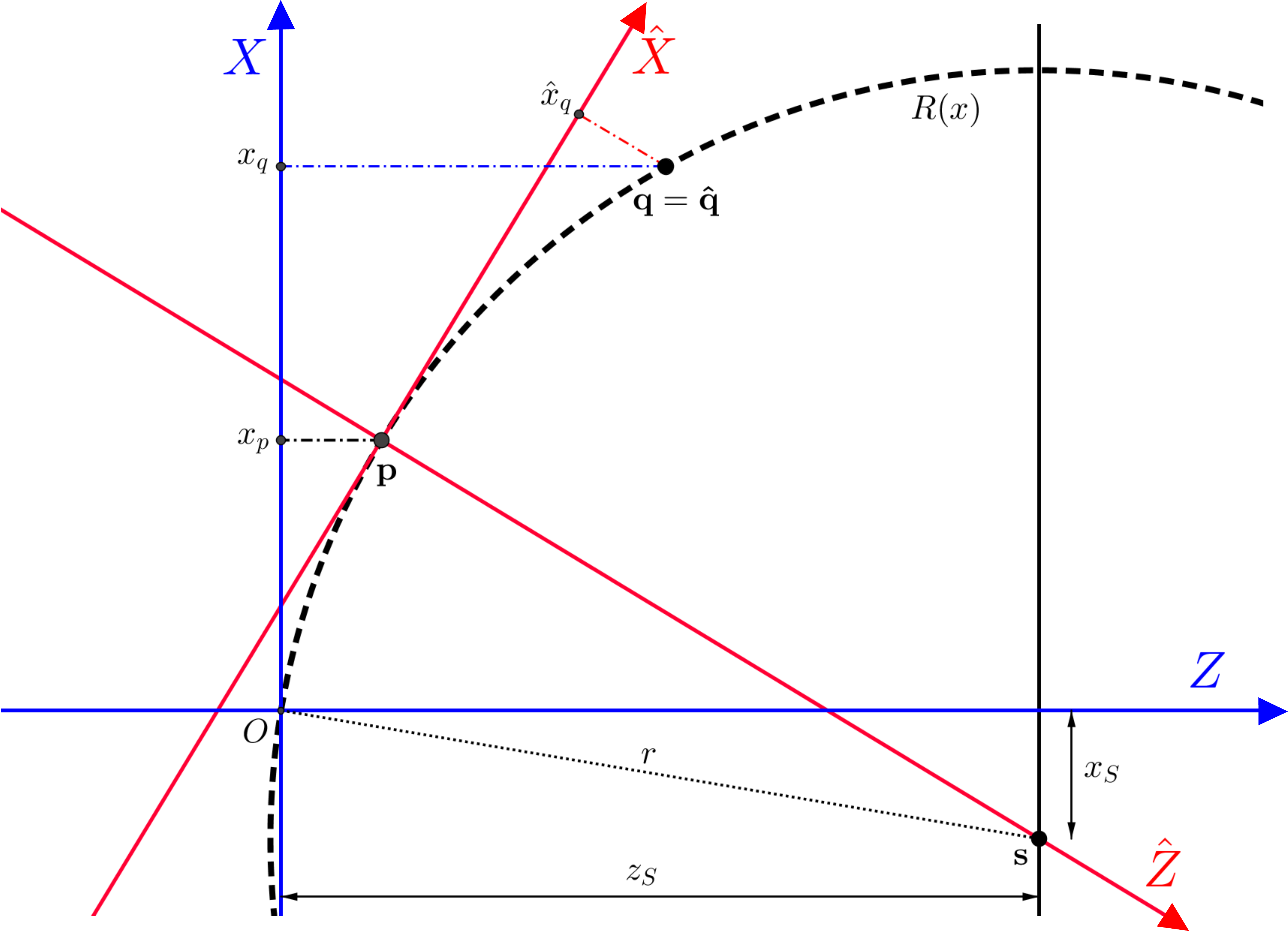}}
\caption{(Color online) Transfer of function values between coordinate
  systems: The center point for the transformation is $\p$ and
  the point $\q$ is an arbitrary point on the reference
  sphere. A function $f$ defined on the reference sphere whose value is given
  in exit pupil coordinates by $f(\x_q)$ must return the same value for
  the local coordinates $\hat{f}(\xh_q)$, i.e. $f(\x_q) =
  \hat{f}(\xh_q)$.
 }
\label{fig:coordinates_function}
\end{figure}

In a local coordinate system, the point $\q = \Tpinv \qh$ has a
different set of coordinates $\qh$. However, the function values of
the OPD function $\Wh(\xh_q,\yh_q)=\W(\x_q,\y_q)$ in both systems must
be the same since $\q$ and $\qh$ are only different coordinates of the
same point, see Fig.~\ref{fig:coordinates_function}. We therefore
define

\begin{equation}
  \Wh(\xh,\yh) := W( \x(\xh,\yh), \y(\xh,\yh) ), 
  \label{eq:deflocalOPD}
\end{equation}

where the functions

\reply{
\begin{eqnarray}
  &&\left(
  \begin{array}{c}
    x(\xh,\yh; \xp,\yp)\\
    y(\xh,\yh; \xp,\yp)\\
    z(\xh,\yh; \xp,\yp)\\
    1
  \end{array}
  \right)
  =
  \Tpinv
  \left(
  \begin{array}{c}
    \xh \\
    \yh \\
    \Rh(\xh,\yh)\\
    1\\
  \end{array}
  \right), 
  \label{eq:coordtransf2D}
\end{eqnarray}
}

perform a remapping of the two different parameterizations of the
reference sphere. For clarity, we have explicitly denoted the
dependence on the evaluation point $\p$. For
the transfer of functions between the two systems, only the functions
$\x(\xh,\yh)$ and $\y(\xh,\yh)$ are significant.

The construction above serves as a general tool to transfer all quantities
of interest into a local coordinate system, i.e. in addition to the
OPD function $\W(x,y)$, the wavefront $\wavefront(x,y)$ and the
normalized ray direction of the aberrated ray $\n(x,y)$ are defined locally via

\begin{eqnarray}
  \phih(\xh, \yh) &:=& \wavefront(x(\xh,\yh),y(\xh,\yh)),  \reply{\,\, \textrm{and}} \nonumber \\
  \nh(\xh, \yh)   &:=& \n(x(\xh,\yh),y(\xh,\yh)),
\end{eqnarray}

by linking them to their global definition in exit pupil coordinates.

\begin{figure}[t]
\centering
\fbox{\includegraphics[width=0.95\linewidth]{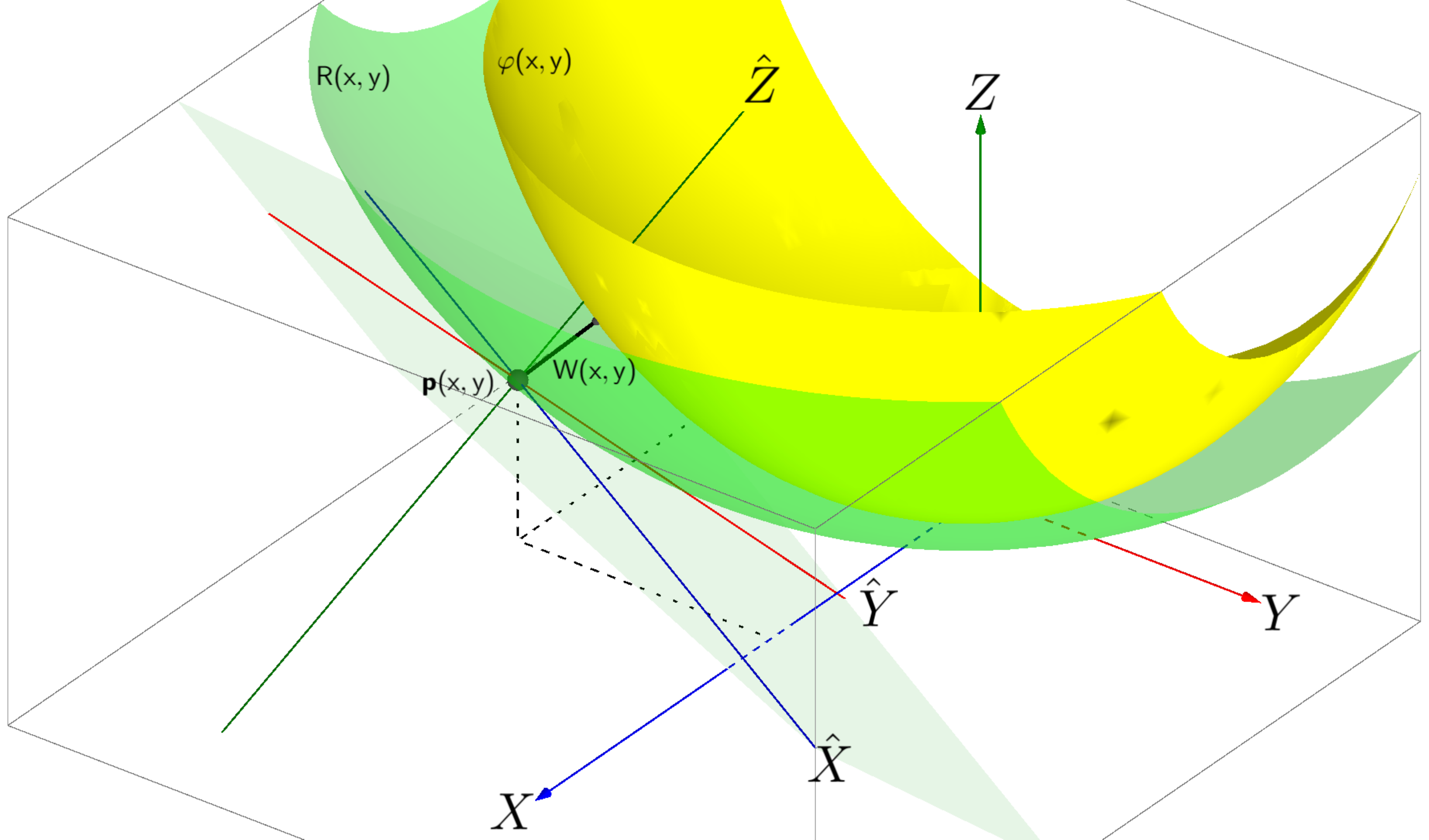}}
\caption{(Color online) Geometry considered. $XYZ$ are the global
  coordinates. $\hat{X}\hat{Y}\hat{Z}$ are the local coordinate axes
  at point $\p$. The wavefront $\W(x,y)$ and the reference sphere
  $\R(x,y)$ are plotted along with the tangent space for the point
  $\p$.}
\label{fig:wavefront2D_coord_change}
\end{figure}

\section{Exact 2D calculation of the wavefront from the wave aberration}
\label{sec:2D_derivation}

\subsection{Constructing a Tangent Plane to the Wavefront}

The derivation for the 2D case utilizes the 1D result of
Sect.~\ref{sec:1D_derivation} and proceeds along similar lines. To
restate the problem: knowing the reference sphere $\R(x,y)$ and the
OPD function $\W(x,y)$ we want to calculate the corresponding
wavefront $\wavefront(x,y) : \fieldR^2 \mapsto \fieldR^3$, which is
now a function that takes a 2D coordinate to a point in three-dimensional space.
According to the wavefront and OPD definitions, 

\begin{equation}
  \wavefront(\xp,\yp) = \p(\xp,\yp) + \W(\xp,\yp)\n(\xp,\yp),
  \label{eq:wavefront_definition_2D}
\end{equation}
where $\n(\xp,\yp)$ is again the unknown normalized direction of the
aberrated ray and $\p(\xp,\yp) = ( \xp, \yp, R(\xp,\yp) )^T$ represents
the evaluation point on the reference sphere. The derivation is
similar to the 1D case, namely the combination of a coordinate
transformation and the calculation of a limit. The geometry of the
setting as well as the global and local coordinate systems involved are
illustrated in Fig.~\ref{fig:wavefront2D_coord_change}.

We continue the derivation in local coordinates. We now study two
spheres with their centers located on the surface of the reference
sphere. Both spheres are tangent to the wavefront. One of them is
fixed at the origin of the local coordinate system while the other one
approaches it. Ultimately, the two spheres coincide at the
limit. Since the second sphere can be approaching from an arbitrary
direction, we instead consider two pairs of spheres moving along the
local coordinate axes $\hat{X}$ and $\hat{Y}$, respectively. In doing
so, we obtain equations where the $\hat{X}$ and $\hat{Y}$ components
are independent. To each pair of spheres, there is a tangent cone with
its vertex located in the $\hat{X}\hat{Z}$ and $\hat{Y}\hat{Z}$ planes
respectively.

\begin{figure}
\centering
\fbox{\includegraphics[width=0.95\linewidth]{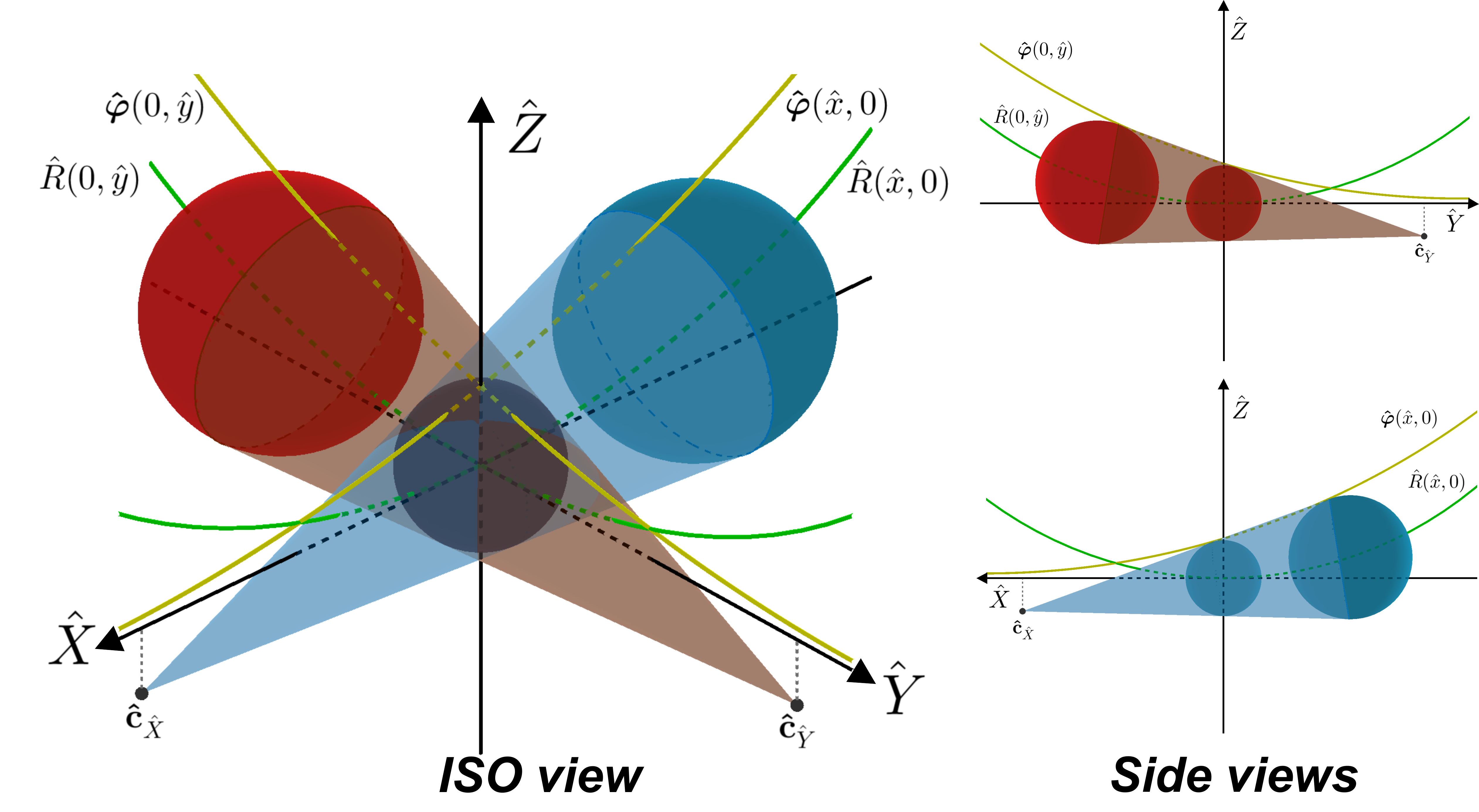}}
\caption{(Color online) Approaching spheres. The wavefront surface and
  the reference sphere are represented as single profiles to simplify
  the graphics. The two \reply{approaching} spheres are centered on the reference
  sphere and their tangent cones have vertices $\chxh$
  and $\chyh$.}
\label{fig:approaching_spheres_solid}
\end{figure}

The two pairs of spheres are illustrated in
Fig.~\ref{fig:approaching_spheres_solid}. In this Figure, the
wavefront surface and the reference sphere previously plotted in
Fig.~\ref{fig:wavefront2D_coord_change}, are now replaced by single
profiles. In the side views we see the two pairs of spheres. The
objects tangent to each are now cones with vertices $\chxh$ and
$\chyh$. Employing the same analysis as in
Fig.~\ref{fig:Tangents_coincide}, we find the equations for the
coordinates of the vertices.

\begin{figure}
\centering
\fbox{\includegraphics[width=0.95\linewidth]{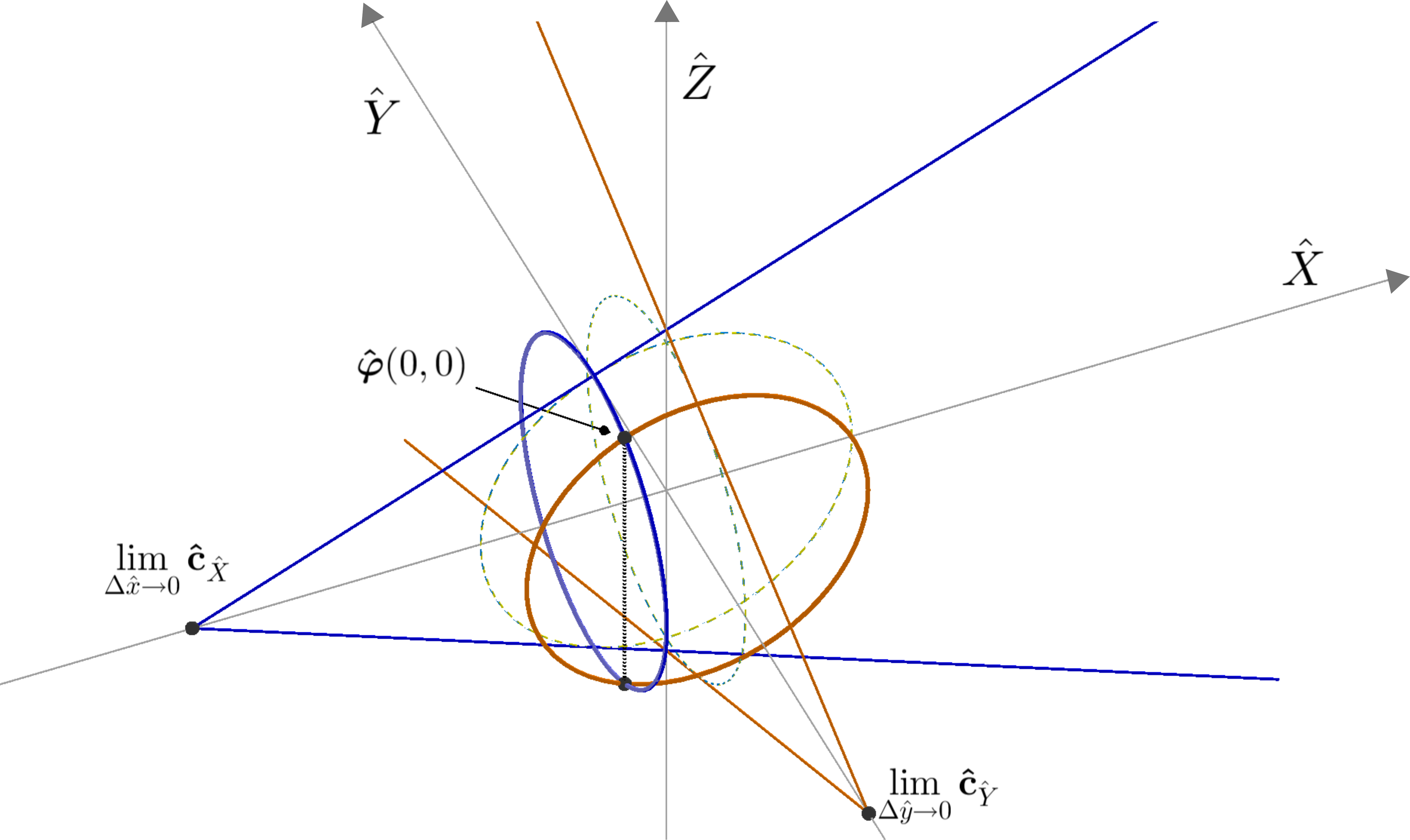}}
\caption{(Color online) Coinciding spheres. The vertices are now
  located along the corresponding axis. The tangent circles intersect
  at two points, for two solutions of the wavefront
  $\hat{\wavefront}(0,0)$. We select the solution whose sign fits the
  OPD.}
\label{fig:approaching_spheres_lines_coincide}
\end{figure}

Consequently, when the spheres are coinciding at the limit, we find
the coordinates for the vertices of the cones to be

\begin{equation}
  \lim_{\Delta\xh \to\\0} \chxh =
  \left(
  \begin{array}{c}
    - \Wh(0,0) \bigg/ \frac{\partial\Wh}{\partial\xh}\big|_{(0,0)} \\
    0 \\
    0
  \end{array}
  \right),
  \label{eq:limit_Cx}
\end{equation}

and

\begin{equation}
  \lim_{\Delta\yh\to\\0} \chyh =
  \left(
  \begin{array}{c}
    0 \\
    -\Wh(0,0) \bigg/ \frac{\partial\Wh}{\partial\yh}\big|_{(0,0)}\\
    0
  \end{array}
  \right),
  \label{eq:limit_Cy}
\end{equation}

where $\Delta\xh$ and $\Delta\yh$ represent the distances between
centers, along the corresponding axis, for each pair of approaching
spheres. The coinciding spheres are shown in
Fig.~\ref{fig:approaching_spheres_lines_coincide}.  The limit points
$\chxh$ and $\chyh$ are located on the axes $\hat{X}$ and $\hat{Y}$
respectively. The tangential circles are located in planes parallel to
the $\hat{X}\hat{Z}$ and $\hat{Y}\hat{Z}$ planes.

In the case that either limit approaches infinity, the solution
reduces to the 1D case. If both limits are infinite, the wavefront
point is at a distance $\hat{W}(0,0)$ along the $\hat{Z}$ axis.

Similar to the 1D case where we obtained the wavefront point $\Wh(0)$
using Eqs.~(\ref{eq:xz_bar}), we compute the wavefront point
$\Wh(0,0)$ as the point of intersection of the limiting tangent
circles, as shown in
Fig.~\ref{fig:approaching_spheres_lines_coincide}. The $\hat{X}$ and
$\hat{Y}$ coordinates are obtained using triangle relations as in
Fig.~\ref{fig:Tangents_coincide}. The $\hat{Z}$ component is
determined by observing that both circles in
Fig.~\ref{fig:approaching_spheres_lines_coincide} are part of the
limit sphere. The wavefront in the local coordinate system is
therefore given by

\begin{equation}
  \phih(0,0) = \Wh(0,0)
  \left(
  \begin{array}{c}
    -\frac{\partial\Wh}{\partial\xh}\big|_{(0,0)} \\ 
    -\frac{\partial\Wh}{\partial\yh}\big|_{(0,0)} \\
    \sqrt{ 1 - \left( \frac{\partial\Wh}{\partial\xh}|_{(0,0)} \right)^{2}
      - \left( \frac{\partial\Wh}{\partial\yh}|_{(0,0)} \right)^{2} }
  \end{array}
  \right).
\label{eq:wavefront_derivation_2D}
\end{equation}

The normalized ray direction in local coordinates is the vectorial
part of the wavefront point and is given by

\begin{equation}
  \nh(0,0) =
  \left(
  \begin{array}{c}
    -\frac{\partial\Wh}{\partial\xh}\big|_{(0,0)} \\ 
    -\frac{\partial\Wh}{\partial\yh}\big|_{(0,0)} \\
    \sqrt{ 1 - \left( \frac{\partial\Wh}{\partial\xh}|_{(0,0)} \right)^{2}
      - \left( \frac{\partial\Wh}{\partial\yh}|_{(0,0)} \right)^{2} }
  \end{array}
  \right).
\label{eq:raydir_derivation_2D}
\end{equation}

\subsection{Explicit Expressions in Global Coordinates}

In the previous sections we have derived equations for the wavefront
point and the aberrated ray direction in local coordinates. We now
relate them back to the global coordinate system. 

\reply{Since both the parameters and the values} of the functions in
Eqs.~(\ref{eq:wavefront_derivation_2D})
and~(\ref{eq:raydir_derivation_2D}) are in local coordinates, two steps
are necessary:

\begin{enumerate}
  \item a replacement of local wavefront and wavefront derivative
    terms by their corresponding expressions in global coordinates,
    followed by
  \item a back-transformation (using Eq.~(\ref{eq:coordtransf2D})) of
    the {\em still local} result into global coordinates.
\end{enumerate}

Before carrying out the two steps, we note that the local OPD function
$\Wh(0,0)$ evaluated at the local origin is equal to the global OPD
function $\W(\xp,\yp)$ evaluated at the exit pupil coordinates of
point $\p$ by Eq.~(\ref{eq:coordtransf2D}).

Second, the 2D expression for the local wavefront derivatives in
global coordinates is

\begin{equation}
  \left(
  \begin{array}{c}
    \frac{\partial \Wh}{\partial \xh}|_{(0,0)}\\
    \frac{\partial \Wh}{\partial \yh}|_{(0,0)}\\    
  \end{array}
  \right)
  =
  \gradWhat|_{(0,0)} = \Tbar \gradW|_{(\xp,\yp)}, 
  \label{eq:locgradglob2D}
\end{equation}

where

\begin{equation}
  \Tbar =
  \left(
  \begin{array}{cc}
    \Tpinvel{11} & \Tpinvel{21} \\
    \Tpinvel{12} & \Tpinvel{22} \\
  \end{array}
  \right),
\end{equation}

\reply{with  $\Tpinvel{11},\ \Tpinvel{21},\ \Tpinvel{12},\  \Tpinvel{22}$ being
the elements of the upper left $2\times2$ sub-matrix from Eq.~(\ref{eq:Tpinv2D})}.
The relation can be found by differentiating the definition of the
function $\Wh$, Eq.~(\ref{eq:deflocalOPD})\reply{;} the Jacobian of the coordinate transformation,
Eq.~(\ref{eq:coordtransf2D}), is found to be $ \Tbar$ and
Eq.~(\ref{eq:locgradglob2D}) follows by the chain rule. Note that
$\Tbar$ is the transpose of the upper left $2\times2$ sub-matrix of
$\Tpinv$ and therefore also dependent on $\p$.

With these prerequisites we can now perform the transformation in two
steps.

\paragraph{Wavefront Point:}

Performing step 1 yields

\begin{eqnarray}
  \phih(0,0) &=&
  \left( \begin{array}{c}
    -\hat{W}(0,0)\gradWhat|_{(0,0)}\\
    \hat{W}(0,0) \cdot \sqrt{ 1 - \gradWhat^T\cdot\gradWhat|_{(0,0)}}\\
    1
  \end{array}\right) \nonumber \\
  &=& 
  \left( \begin{array}{c}
    -W(\xp,\yp) \Tbar\gradW|_{(\xp,\yp)}\\
     W(\xp,\yp) \sqrt{ 1 - \gradW^T  \displaystyle\Tbar^T \cdot \Tbar\gradW|_{(\xp,\yp)}}\\
     1
  \end{array}\right). \nonumber \\
\end{eqnarray}

Defining

\begin{equation}
  n_f= ( 1 - \gradW^T \displaystyle{\Tbar^T}\cdot\Tbar\gradW|_{(\xp,\yp)})^{1/2}
\end{equation}

and performing step 2 yields

\begin{eqnarray}
  \wavefront(\xp,\yp) &=& \Tpinv \phih(0,0) =
  \Tpinv
  \left(
  \begin{array}{c}
    -\W \cdot\Tbar \gradW\\ \W \cdot\nf\\
    1
  \end{array}
  \right).
  \label{eq:wfglob2D}
\end{eqnarray}

which is the 3D wavefront point in global (exit pupil) coordinates as
a function of the exit pupil coordinates. Note that the $X$- and
$Y$-components of $\wavefront(\xp,\yp)$ will typically not be $(\xp,\yp)$
unless $\W(\xp,\yp)=0$, Eq.~(\ref{eq:coordtransf2D}).

\paragraph{Ray Direction:}

The aberrated ray direction is similarly transformed into global
coordinates. Step 1 results in

\begin{equation}
  \nh(0,0) =
  \left(
  \begin{array}{c}
    - \Tbar \gradW|_{(\xp,\yp)} \\
      n_f \\
      0
  \end{array}\right).
  \label{eq:normalloc2D}
\end{equation}

Applying step 2, we obtain

\begin{equation}
  \n(\xp,\yp) =
  \Tpinv
  \left( \begin{array}{c}
    - \Tbar \gradW|_{(\xp,\yp)} \\
    n_f \\
    0
  \end{array}\right).
  \label{eq:normalglob2D}
\end{equation}

Since only three components of the homogeneous direction vector $\n$
are non-zero, only the upper left $3\times3$ matrix of $\Tpinv$ is
effective. For general transformations $M$, normals need to be
transformed via $M^{-T}$~\cite{glassner1987transformation}. In our case, the upper left $3\times3$
matrix of $\Tpinv$ is orthonormal and therefore its own inverse
transpose.

Eqs.~(\ref{eq:wfglob2D}) and~(\ref{eq:normalglob2D}) are the key
results of this paper. \reply{An analytical proof that the ray
  directions $\n$ are orthogonal to the wavefront $\wavefront$ at
  every point and that the wavefront is located at the OPD distance of
  $W$ is given in the Appendix.}

\section{Connection to the ray aberrations}

\label{sec:connection_ray_aberration}

Once the wavefront point and the aberrated ray direction are known,
they can be used to compute the transverse ray aberrations. We make a
connection to the classical approximation, Eqs.~\ref{eq:classic_eq},
that relates wave and ray aberrations and derive the exact
conditions for the approximation to be valid.

The transverse ray aberrations $\boldsymbol{\epsilon}$ are obtained by computing
the aberrated image point $\i$ and subtracting the paraxial image
point $\s=(x_s,y_s,z_s)^T$ from it, i.e.

\begin{equation}
  \boldsymbol{\epsilon} = \i - \s.
  \label{eq:traeasy}
\end{equation}

The aberrated image point $\i$ is obtained by computing the ray
intersection of the aberrated ray with the image plane situated at
$z_s$. The aberrated ray passes, by definition, through the wavefront
point $\wavefront(\xp,\yp)$, Eq.~(\ref{eq:wfglob2D}), and has the direction
$\n(\xp,\yp)$, Eq.~(\ref{eq:normalglob2D}). We compute the path length
$\opdtoip$ to the image plane by

\begin{equation}
  \opdtoip = \frac{z_s - [\wavefront]_z}{[\n]_z} = \frac{(z_s-\R(\xp,\yp))}{[\n]_z} - \W(\xp,\yp)
\end{equation}

where $([\wavefront]_x,[\wavefront]_y,[\wavefront]_z)^T$ are the
components of $\wavefront(\xp,\yp)$, $([\n]_x,[\n]_y,[\n]_z)^T$ the
components of $\n(\xp,\yp)$, and Eq.~(\ref{eq:wavefront_definition_2D}) has been
used. The aberrated image point can now be written as

\begin{eqnarray}
  \i &=& \wavefront(\xp,\yp) + \opdtoip\n(\xp,\yp) \nonumber \\
   &=& \p + \W(\xp,\yp)\n(\xp,\yp) + \opdtoip\n(\xp,\yp) \nonumber \\
  &=& \p + \frac{(z_s-\R(\xp,\yp))}{[\n]_z} \n(\xp,\yp). \nonumber \\
  &=&
  \left(
  \begin{array}{c}
    \xp + (z_s - \R(\xp,\yp))\displaystyle{\frac{[\n]_x}{[\n]_z}}\\
    \yp + (z_s - \R(\xp,\yp))\displaystyle{\frac{[\n]_y}{[\n]_z}}\\
    z_s
  \end{array}
  \right).
\end{eqnarray}

We see that the point $\i$ is indeed in the image plane. Using
Eq.~(\ref{eq:traeasy}) and ignoring the zero z-component, the transverse
ray aberrations become

\begin{eqnarray}
  \epsilon_x =  (\xp-\xs) + (z_s - R(\xp,\yp))\frac{[\n]_x}{[\n]_z} \nonumber \\
  \epsilon_y =  (\yp-\ys) + (z_s - R(\xp,\yp))\frac{[\n]_y}{[\n]_z}
\end{eqnarray}

which, with the help of a computer algebra package, can be simplified
to a closed form in terms of the OPD derivative and the paraxial image
position $\s$:

\renewcommand{\dWdx}{\frac{\partial\W}{\partial \x}}
\renewcommand{\dWdy}{\frac{\partial\W}{\partial \y}}

\begin{eqnarray}
  &&\epsilon_x =  -\frac{r^2 \dWdx} {A-B} \nonumber \\
  &&\epsilon_y =  -\frac{r^2 \dWdy} {A-B} \nonumber \\
  \label{eq:exacttra}
\end{eqnarray}

with 

\begin{eqnarray}
  B &=&\dWdx\Delta x+\dWdy\Delta y,  \nonumber \\
  A &=&\sqrt{ r^2\left(1-\left(\dWdx\right)^2-\left(\dWdy\right)^2\right) + B^2}
\end{eqnarray}

and $\Delta x=(\xp-\xs)$ and $\Delta
y=(\yp-\ys)$. Eqs.~(\ref{eq:exacttra}) are the exact equations of the
transverse ray aberrations in terms of the OPD function.
We now show that the classic approximation can be obtained as a special case.

Taking the limit as $\Delta x \to 0, \Delta y \to 0$ results in

\begin{eqnarray}
  \lim_{\Delta x \to 0, \Delta y \to 0} \epsilon_x &=& -\frac{r \dWdx} {\sqrt{   \left(1 - \left(\dWdx\right)^2-\left(\dWdy\right)^2\right) }} \nonumber \\
  \lim_{\Delta x \to 0, \Delta y \to 0} \epsilon_y &=& -\frac{r \dWdy} {\sqrt{   \left(1 - \left(\dWdx\right)^2-\left(\dWdy\right)^2\right) }}, 
\end{eqnarray}

from which we obtain the classical form, Eqs.~(\ref{eq:classic_eq}), in
the case of $||\gradW|| \ll 1$. The conditions for the validity of the
classical approximation are therefore:

\begin{enumerate}[label=(\roman*)]
  \item the evaluation position $(\xp,\yp)$ in the exit pupil approaches the
    paraxial image coordinate $(\xs,\ys)$, and
  \item the gradient norm of the OPD  $||\gradW|| \ll 1$.
\end{enumerate}

It is interesting to note that the best approximation is not obtained
in the origin of the exit pupil, but in the exit pupil position
closest to the paraxial image.

\section{Examples of wavefront and ray aberration calculations}
\label{sec:examples}

In the following, we provide three representative examples of
computations with our analytic expressions.
We provide a comparison with results obtained from the classical
approximation in Eqs.~(\ref{eq:classic_eq}). The comparisons are
performed for both the transverse ray aberrations and the
reconstructed wavefront. The error between our computation and the
classical approximation is studied for all three examples.

\begin{figure*}[htbp]
\centering
\fbox{\includegraphics[width=\textwidth]{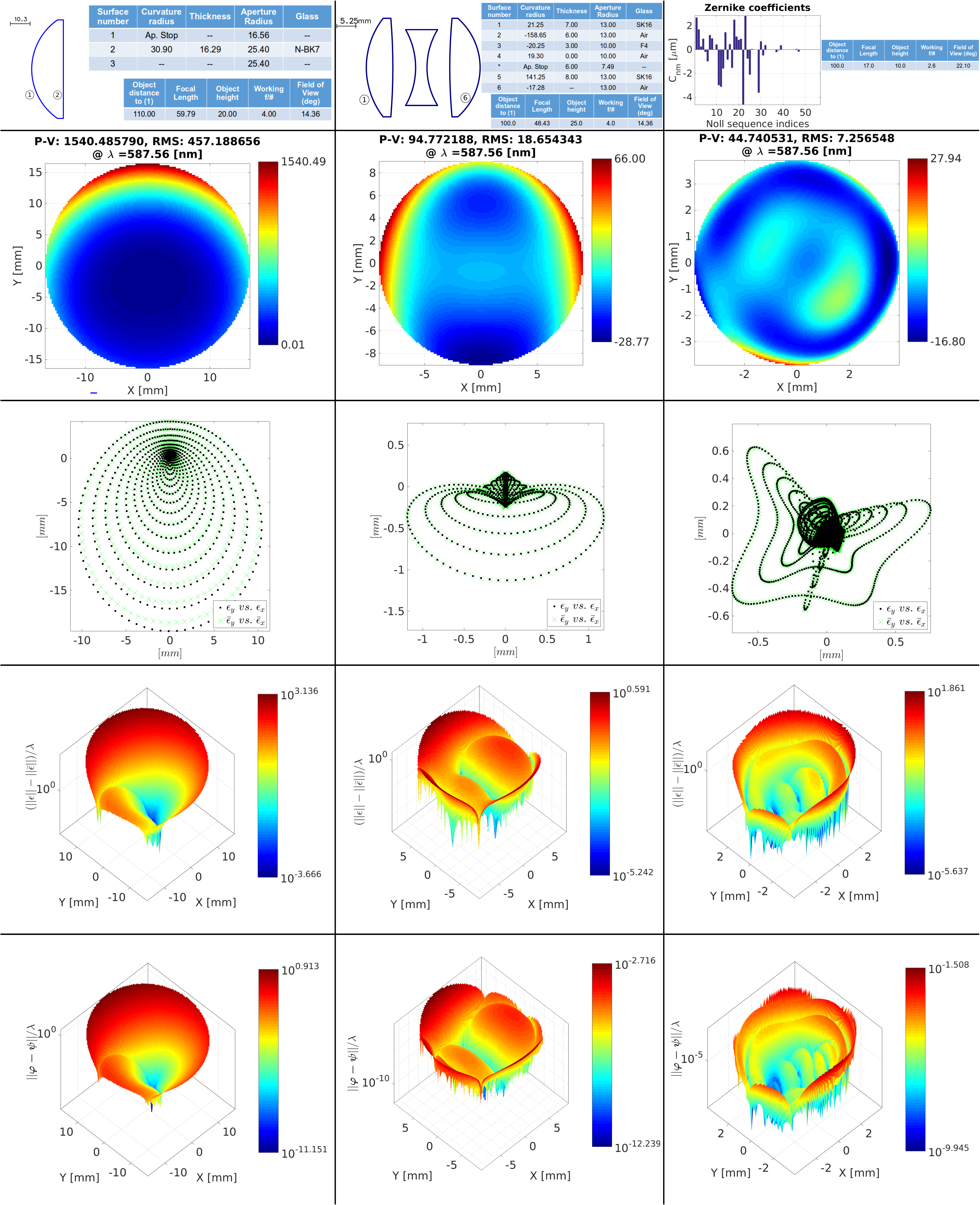}}
\caption{(Color online) Multiple computations for a singlet, a Cooke triplet and a post-surgery corneal aberration map.}
\label{fig:Results}
\end{figure*}

The results are summarized in Fig.~\ref{fig:Results}. Each column is
dedicated to a different test case and each row contains the same type
of plot for each example. First, we briefly describe the information
found in each row and later discuss the results in the context of each
of the optical systems.

The first row shows a scaled system diagram, together with the lens
prescription and the imaging conditions, reporting all quantities in
millimeters. We obtain the first order system properties and a Zernike
decomposition of the OPDs from OSLO, a commonly known Optics design
software. \reply{The number of polynomials for the fit are set to the 
maximum available, yielding 37 Zernike coefficients.}  We use the Zernike
expansions as ground truth OPD functions, shown in the second
row. Additionally, we provide the \textit{peak-to-valley} (P-V) and
\textit{RMS} values to emphasize the degree of the aberrations.

In the third row we plot the transverse ray aberrations for both
directions, which can be considered as a re-centered spot diagram in
the paraxial image plane. We simultaneously plot the exact ray
aberrations $\epsilon_y$ vs. $\epsilon_x$ and the classical ray
aberrations, now re-named to $\bar{\epsilon}_{y}$
vs. $\bar{\epsilon}_{x}$.

The fourth row presents a comparison between the exact ray aberrations
and the classical approximation, by plotting the error in the norms of
the \reply{transverse} ray aberrations $||\boldsymbol{\epsilon}|| - ||\bar{\boldsymbol{\epsilon}}||$,
with $||\boldsymbol{\epsilon}|| = ({\trax}^2 + {\tray}^2)^{1/2}$ being the norm of
the exact \reply{transverse} ray aberrations, Eqs.~(\ref{eq:exacttra}), and
$||\bar{\boldsymbol{\epsilon}}|| = ({\bar{\epsilon}_{x}}^2 +
{\bar{\epsilon}_{y}}^2)^{1/2}$ the norm for the classic equations,
Eqs.~(\ref{eq:classic_eq}). This surface error is scaled in
wavelengths and plotted on a logarithmic $Z$ axis to enhance its
dynamic range.

Finally, the last row presents a comparison of the exactly computed
wavefront, Eq.~(\ref{eq:wfglob2D}), with an approximation obtained from
the classical equations. For computing the latter, we propose the
following procedure: we connect the reference sphere evaluation point
$\p$ with the approximation of the aberrated image point $\bar{\i}$,
obtained by adding the classical transverse ray aberrations
$(\bar{\epsilon}_{x}, \bar{\epsilon}_{y})^T$ to the paraxial image
point $\s$. The direction of this 3D line is an approximation to the
true aberrated ray direction $\n$, Eq.~(\ref{eq:normalglob2D}). We use
it in conjunction with Eq.~(\ref{eq:wavefront_definition_2D}) to
construct an approximated wavefront $\wfa$. The difference in 3D
positions $||\wavefront(\x,\y)-\wfa(\x,\y)||$ is shown in the last
row, again scaled in wavelengths and with a logarithmic $Z$ axis. We
include this additional comparison, since the wavefront shape is
important for an exact computation of wave-optical point spread
functions using Huygens' principle. We now discuss the individual test cases.

\paragraph{(i) Singlet:}
This system is one of the simplest optical set-ups with
correspondingly large aberration values, especially at lower
f-numbers. We chose a working $f / \# = 4$, focal length $f \approx
60\mm$ at a wavelength $\lambda = 587.56\nm$. In the OPD plot, a
large astigmatic component can be identified, along with a large P-V
value.  The spot diagram indicates that the classical approximation has a
large error for the most external intersection points. The ray aberration
error plot further emphasizes this fact, presenting a maximum in the
range of $10^{3}\lambda$. The corresponding wavefront error has a
maximum value on the order of $8\lambda$.

\paragraph{(ii) Cooke Triplet:}
We continue, by studying a more complex and well known system: the
Cooke Triplet. The lens prescription is given in the plot. The imaging
conditions closely resemble the singlet example, with a working $f /
\# = 4$, focal length $f \approx 50\mm$ at a wavelength $\lambda =
587.56\nm$. The OPD plot indicates a better corrected system, with
again a large astigmatic component. The spot diagram indicates that
the intersection points from the classical approximation closely
resemble the exact calculations. The errors of the classical
approximation in both, the ray aberration and the wavefront, are now
significantly smaller, with maximum values in the range of $7\lambda$
and $2\times10^{-3}\lambda$\reply{,} respectively.

\paragraph{(iii) Post-Surgery Cornea:}
We select this example as a test case presenting aberrations of higher
order, in contrast to the previous examples where the OPD functions
were slowly varying. The Zernike coefficients are extracted from a
post-LASIK-surgery corneal topography~\cite{Fricker2008}. In the first
row of Fig.~\ref{fig:Results}, we present the list of Zernike
coefficients, measured in microns, were the values are listed using
the Noll index ordering~\cite{Noll1976}. To maintain the context of an
eye as the optical system, we select a working $f / \# = 2.6$, focal
length $f = 17\mm$ at a wavelength $\lambda = 587.56\nm$. The OPD plot
shows a more complex structure giving rise to an unusual spot
diagram. The maximum error of the classical approximation in the ray
aberrations is now in the order of $70\lambda$ and the corresponding
error for the wavefront comparison has a maximum value on the order of
$3\times10^{-2}\lambda$.

Comparing the errors between the cornea and the Cooke triplet
examples, we observe the dependence of the approximation error of the
classical Eqs.~(\ref{eq:classic_eq}) on the derivatives of the wave
aberration function. As stated before, Eq.~(\ref{eq:exacttra}), the
classical approximation becomes more accurate for smaller gradients of
the OPD function. In the post-surgery cornea example, even though the
magnitude of the aberrations is smaller than in the Cooke triplet
case, we observe approximation errors that are an order of magnitude
larger.

\section{Conclusions}
\label{sec:conclusions}

We have derived exact analytic expressions for the wavefront surface,
the aberrated ray directions, and the transverse ray aberrations for
the standard definition of the optical path difference as a phase
delay along the aberrated ray.

A transition to the local tangent frames of the reference sphere
enables a Huygens-like geometric construction of the wavefront as an
envelope of spheres and yields additional constraints on the OPD
function. \reply{We show in the Appendix,} that every OPD function satisfying the
constraints has an associated wavefront and that the constructed
wavefronts and rays fulfill the wavefront properties exactly.

The exact aberrated rays yield exact equations for the transverse ray
aberrations. We have identified the precise conditions for the
classical approximation to hold. The conditions differ from commonly
held assumptions. The relevant factors are small OPD gradients, as
opposed to the OPD magnitude, in conjunction with evaluation positions
close to the paraxial image coordinates, as opposed to the pupil
center.

We presented numerical simulations to illustrate the errors arising
from the classical approximation for typical scenarios. The
simulations provide a quantitative background for the theoretical
results.

\section*{Acknowledgments}

We thank the anonymous reviewers for their helpful advice and
attention to detail.  This work has been supported by the German
Research Foundation (DFG) through Emmy-Noether grant IH 114/1-1.

\appendix

\renewcommand{\dWdx}{\frac{\partial \W}{\partial \xp}}
\renewcommand{\dWdy}{\frac{\partial \W}{\partial \yp}}

\reply{\section*{Appendix: Validating the Wavefront Properties}}
\label{app:validation}

\reply{We now validate the wavefront properties of the wavefront derived in Eq.~(\ref{eq:wfglob2D}) and the ray direction from Eq.~(\ref{eq:normalglob2D}). Two conditions must be met for these quantities to be compatible:}

\begin{enumerate}[label=(\roman*)]
  \item The wavefront point $\wavefront(\xp,\yp)$ obtained via
    Eq.~(\ref{eq:wfglob2D}) must have a distance of $\W(\xp,\yp)$ from
    the point $\p$, and
   \item The aberrated ray direction $\n(\xp,\yp)$,
     Eq.~(\ref{eq:normalglob2D}), must be orthogonal to the wavefront in
     the point $\wavefront(\xp,\yp)$.
\end{enumerate}

The two conditions constitute the definition of a wavefront. In
showing that they are met by the quantities in Eqs.~(\ref{eq:wfglob2D})
and~(\ref{eq:normalglob2D}), we prove their correctness. In addition,
the proof also shows that every OPD function, according to the
definition of Sect.~\ref{sec:alternative_definitions}, has a
corresponding wavefront.

\paragraph{(i) Wavefront Distance:}

First, from Eqs.~(\ref{eq:wfglob2D}) and~(\ref{eq:normalglob2D}), we
\reply{verify} that Eq.~(\ref{eq:wavefront_definition_2D}) holds.

We need to show that $||\n(\xp,\yp)||=1$. For, in this case
$||\wavefront(\xp,\yp)-\p|| = |W(\xp,\yp)|$. Using Eq.~(\ref{eq:normalloc2D}) in
local coordinates, the result is readily obtained:

\begin{eqnarray}
  ||\n(\xp,\yp)||^2 &=& ||\nh(0,0)||^2 =
  \left(-\Tbar\gradW\right)^2 + n_f^2 \nonumber \\
  &=&
  \gradW^T\displaystyle\Tbar^T \cdot \Tbar\gradW + n_f^2 \nonumber \\
  &=&  \gradW^T\displaystyle\Tbar^T \cdot \Tbar\gradW + 1 - \gradW^T\displaystyle\Tbar^T \cdot \Tbar\gradW  \nonumber \\
                     &=& 1.
  \label{eq:normnlocone2D}
\end{eqnarray}

Since the upper left $3\times3$ sub-matrix of transformation $\Tpinv$
is orthonormal, it does not change the length of the normal vector
when changing to global coordinates.
\begin{flushright}$\qedsymbol$\end{flushright}

\paragraph{(ii) Orthogonality of Wavefront and Aberrated Ray Direction:}

We need to show
\begin{eqnarray}
  \n(\xp,\yp)^T \cdot \frac{\partial \wavefront(\xp,\yp)}{\partial \xp} &=& 0,\,\, \textrm{and} \nonumber \\
  \n(\xp,\yp)^T \cdot \frac{\partial \wavefront(\xp,\yp)}{\partial \yp} &=& 0.
\end{eqnarray}

We proof the equality for the x-tangent vector $\frac{\partial
  \wavefront}{\partial \xp}$, the derivation for the y-tangent vector
being strictly similar. It is important to perform the proof in global
coordinates since the local coordinate system changes when changing
the evaluation position~$\p$. 

Denoting the homogeneous local coordinate origin as $\hat{O} =
(0,0,0,1)^T$, the dot-product between the tangent vector and the
aberrated ray direction is given by

\renewcommand{\homzerovec}{\hat{O}}

\begin{eqnarray}
  \n^T \frac{\partial \wavefront}{\partial \xp} &=& \nh^T \left( \Tpinv \right)^T \Tpinv \frac{\partial }{\partial \xp} \left( \W \nh  + \homzerovec \right) + \nonumber \\
                                             & &    \nh^T \left( \Tpinv \right)^T   \frac{\partial \Tpinv}{\partial \xp} \left( \W  \nh  + \homzerovec \right) \nonumber \\
                                             &=&
  \underbrace{
    \vphantom{\homzerovec} 
    \nh^T \frac{\partial }{\partial \xp} ( \W \nh )}_{(I)\, \textrm{yields}\, \dWdx} + 
  \underbrace{\nh^T \left( \Tpinv \right)^T   \frac{\partial \Tpinv}{\partial \xp} \left( \W  \nh  + \homzerovec  \right)}_{(II)\, \textrm{yields}\, -\dWdx} \nonumber \\
                                                 &=& 0.
  \label{eq:dotproduct2D}
\end{eqnarray}

Simplification of the complete Eq.~(\ref{eq:dotproduct2D}) with a
computer algebra package is, unfortunately, not tractable. As
indicated above, we decompose the equation into term (I) and term
(II). Term (I) will be shown to equal $\dWdx$ ($\dWdy$ for the
y-tangent). Term (II) can be shown to equal $-\dWdx$ ($-\dWdy$ for the
y-tangent) with the help of a computer algebra software.

\paragraph{Term $(I)$:} contributes the majority of cross-terms when multiplying out
Eq.~(\ref{eq:dotproduct2D}). It can be conveniently treated in a
vectorial fashion:

\begin{eqnarray}
  \nh^T \frac{\partial }{\partial \xp} ( \W \nh )
  &=& \nh^T \cdot ( \dWdx \nh + \W \frac{\partial\nh}{\partial \xp} ) \nonumber \\
  &=& \nh^T \nh \cdot \dWdx + \W \nh^T \frac{\partial\nh}{\partial \xp}. \nonumber \\
  &=& \dWdx.
\end{eqnarray}

The last equality is due to the fact that $\nh^T
\nh=1$, as shown in Eq.~(\ref{eq:normnlocone2D}), and

\begin{eqnarray}
  \nh^T \frac{\partial \nh}{\partial \xp}
  &=& \gradW^T\Tbartransp\dTbardx\gradW + \nonumber \\
  & & \gradW^T\Tbartransp \Tbar
      \left(
      \begin{array}{c}
        \frac{\partial^2 W }{\partial \xp^2}\\
        \frac{\partial^2 W }{\partial \yp \partial\xp}
      \end{array}
      \right)
      + \nf \dnfdx \nonumber \\
      &=& 0. 
\end{eqnarray}

The latter equality is due to

\begin{eqnarray}
  \dnfdx &=& - \frac{ \gradW^T\Tbartransp\dTbardx\gradW + 
      \gradW^T\Tbartransp \Tbar
      \left(
      \begin{array}{c}
        \frac{\partial^2 W }{\partial \xp^2}\\
        \frac{\partial^2 W }{\partial \yp \partial\xp}
      \end{array}
      \right) }{n_f}, \nonumber \\
\end{eqnarray}

which can be verified by multiplying out the expressions.

\paragraph{Term $(II)$:} is tractable with computer algebra software. 
Defining

\begin{equation}
  \vec{r}_x = \frac{\partial \Tpinv}{\partial \xp}\phih(0,0), 
\end{equation}

we find that

\begin{equation}
  \n^T \cdot \vec{r}_x = -\dWdx.
\end{equation}

\begin{flushright}
$\qedsymbol$
\end{flushright}

Summarizing, we have shown that the wavefront point and the
aberrated ray directions derived in Sect.~\ref{sec:2D_derivation} are
compatible with the wavefront and the OPD properties. Since the proof has
been constructive, we have shown that every OPD function according to
the definition of Sect.~\ref{sec:alternative_definitions} has indeed
an associated wavefront.

\bibliography{thebibliography}

\begin{thebibliography}{10}
\newcommand{\enquote}[1]{``#1''}

\bibitem{ob:bornwolf}
M.~Born and E.~Wolf, \emph{Principles of Optics} (Cambridge University Press,
  1999), 7th ed.

\bibitem{ob:welford}
W.~Welford, \emph{Aberrations of the Symmetrical Optical System} (Academic
  Press, 1974).

\bibitem{ob:mahajan}
V.~N. Mahajan, \emph{Optical imaging and aberrations. Part 1. , Ray geometrical
  optics} (SPIE press, Bellingham (Wa.), 1998).

\bibitem{ob:malacara}
D.~Malacara and Z.~Malacara, \emph{Handbook of lens design}, Optical
  engineering (Marcel Dekker, New York, 1994).

\bibitem{ob:wyantcreath}
J.~C. Wyant and K.~Creath, \emph{Basic Wavefront Aberration Theory for Optical
  Metrology}, Applied Optics and Optical Engineering, Volume XI (Academic
  Press, 1992).

\bibitem{oa:platt2001}
B.~C. Platt and R.~Shack, \enquote{{History and principles of Shack-Hartmann
  wavefront sensing.}} Journal of Refractive Surgery \textbf{17} (2001).

\bibitem{Neal2002}
D.~R. Neal, J.~Copland, and D.~A. Neal, \enquote{Shack-hartmann wavefront
  sensor precision and accuracy,} in \enquote{International Symposium on
  Optical Science and Technology,}  (International Society for Optics and
  Photonics, 2002), pp. 148--160.

\bibitem{Mejia2012726}
Y.~Mejia, \enquote{Exact relations between wave aberration and the sagitta
  difference, and between ray aberration and the slope difference,} Optik -
  International Journal for Light and Electron Optics \textbf{123}, 726 -- 730
  (2012).

\bibitem{oa:rayces1964}
J.~Rayces, \enquote{Exact relation between wave aberration and ray aberration,}
  Optica Acta: International Journal of Optics \textbf{11}, 85--88 (1964).

\bibitem{glassner1987transformation}
A.~Glassner and F.~Post, \enquote{On the transformation of surface normals,}
  Tech. rep., Faculty of Mathematics and Informatics, Delft University of
  Technology (1987).

\bibitem{Fricker2008}
P.~Fricker, \enquote{Analyzing lasik optical data using {Z}ernike functions,}
  MATLAB Digest pp. 1--6 (2008).

\bibitem{Noll1976}
R.~J. Noll, \enquote{Zernike polynomials and atmospheric turbulence,} J. Opt.
  Soc. Am. \textbf{66}, 207--211 (1976).

\end{thebibliography}

\end{document}